\documentclass[12pt]{article}

\pdfoutput=1
\DeclareFontFamily{T1}{calligra}{}
\DeclareFontShape{T1}{calligra}{m}{n}{<->s*[1.44]callig15}{}
\DeclareMathAlphabet\mathcalligra   {T1}{calligra} {m} {n}
\DeclareMathAlphabet\mathzapf       {T1}{pzc} {mb} {it}
\DeclareMathAlphabet\mathchorus     {T1}{qzc} {m} {n}
\DeclareMathAlphabet\mathrsfso      {U}{rsfso}{m}{n}
\DeclareMathAlphabet\mathfrcal      {T1}{frcursive}{m}{it}
\DeclareFontFamily{T1}{frcursive}{}
\DeclareFontShape{T1}{frcursive}{m}{n}{<->s*[1.44]callig15}{}
\DeclareMathAlphabet\mathfrcal      {T1}{frcursive}{m}{it}

\usepackage{amssymb}
\usepackage{amsmath}
\usepackage{amsthm}
\usepackage{amssymb}
\usepackage{graphicx}
\usepackage[dvipsnames]{xcolor}
\usepackage{soul}
\numberwithin{equation}{section}
\usepackage{array}
\usepackage{mathtools}
\usepackage{dsfont}
\usepackage{mathrsfs}

\usepackage{BOONDOX-uprscr}

\usepackage{tikz}\usetikzlibrary{matrix,fit}\usetikzlibrary{decorations.pathreplacing,arrows.meta}
\usepackage{tikz-cd} 
\usepackage{varwidth}
\usepackage{enumerate}
\usepackage{appendix}
\usepackage{xfrac}
\usepackage{nicefrac}
\usepackage{mathtools,slashed}

\usepackage{geometry}
\usepackage{nicematrix}

\usepackage[
    backend=bibtex,
    style=alphabetic,
maxbibnames=99,
giveninits=true,
minalphanames=1,
maxalphanames=3,url=false
  ]{biblatex}

\bibliography{SUSYGrassmannian}

\usepackage{mathabx}
\usepackage{empheq}

\usepackage{setspace}

\usepackage{slashed}
\usepackage{upgreek}
\usepackage{appendix}

\usepackage{tabstackengine}

\usepackage{wrapfig}
\usepackage[abs]{overpic}

\usepackage{float}

\fixTABwidth{T}

\newdimen\mytextwidth
\newcommand\rem[2][cyan!40!green]{\noindent\nobreak\hfil\penalty1000\hfilneg
\mytextwidth=\linewidth\advance\mytextwidth by 2mm
\begin{tikzpicture}[baseline=-\the\dimexpr\fontdimen22\textfont2\relax]\node[outer sep=0pt,draw=black,fill=#1,fill opacity=1,text opacity=1,rectangle,rounded corners]{\begin{varwidth}{\mytextwidth}\textcolor{white}{#2}\end{varwidth}};
\end{tikzpicture}\allowbreak
}

\newcommand\whiterem[2][white!]{\noindent\nobreak\hfil\penalty1000\hfilneg
\mytextwidth=\linewidth\advance\mytextwidth by 2mm
\begin{tikzpicture}[baseline=-\the\dimexpr\fontdimen22\textfont2\relax]\node[outer sep=0pt,draw=black,fill=#1,fill opacity=1,text opacity=1,rectangle,rounded corners,line width=1.5pt]{\begin{varwidth}{\mytextwidth}\textcolor{black}{#2}\end{varwidth}};
\end{tikzpicture}\allowbreak
}

\makeatletter
\newsavebox{\@brx}
\newcommand{\llangle}[1][]{\savebox{\@brx}{\(\m@th{#1\langle}\)}%
  \mathopen{\copy\@brx\kern-0.5\wd\@brx\usebox{\@brx}}}
\newcommand{\rrangle}[1][]{\savebox{\@brx}{\(\m@th{#1\rangle}\)}%
  \mathclose{\copy\@brx\kern-0.5\wd\@brx\usebox{\@brx}}}
\makeatother

\usepackage{latexsym}

\newcommand{\bea}{\begin{equation}}
\newcommand{\eea}{\end{equation}}
\newcommand{\bear}{\begin{eqnarray}}
\newcommand{\eear}{\end{eqnarray}}
\newcommand{\bearr}{\begin{eqnarray*}}
\newcommand{\eearr}{\end{eqnarray*}}

\newcommand{\dif}{\mathrm{d}}

\usepackage[boxsize=1em]{ytableau}
\ytableausetup{centertableaux}

\usepackage{tikz}
\usepackage{xparse}
\NewDocumentCommand{\xrightarrows}{ O{}O{} }{%
\mathrel{%
\vcenter{\hbox{%
\begin{tikzpicture}
  \node[minimum width=1cm,minimum height=1ex,anchor=south,align=center] (a){\text{\vphantom{hg}#1}\\[0.5ex] \vphantom{hg}#2};
  \draw[<-] ([yshift=0.35ex]a.west) -- ([yshift=0.35ex]a.east);
  \draw[->] ([yshift=-0.35ex]a.west) -- ([yshift=-0.35ex]a.east);
\end{tikzpicture}
}}%
}%
}

\renewbibmacro{in:}{}

\usepackage{mdframed}

\newbibmacro{string+doi}[1]{
  \iffieldundef{doi}{#1}{\href{http://dx.doi.org/\thefield{doi}}{#1}}}
\DeclareFieldFormat{title}{\usebibmacro{string+doi}{\mkbibemph{#1}}}
\DeclareFieldFormat[article]{title}{\usebibmacro{string+doi}{\mkbibquote{#1}}}

\newmdenv[
  topline=false,
  bottomline=false,
  rightline=false,
  linewidth=2pt,
  skipabove=\topsep,
  skipbelow=\topsep
]{siderules}

\newmdenv[
  topline=false,
  bottomline=false,
  linewidth=2pt,
  skipabove=\topsep,
  skipbelow=\topsep
]{siderulesright}

\makeatletter
\renewcommand{\@seccntformat}[1]{\csname the#1\endcsname.\quad}
\makeatother

\usepackage{setspace}
\usepackage{xpatch}

\makeatletter
\renewcommand{\@chap@pppage}{
  \clear@ppage
  \thispagestyle{plain}
  \if@twocolumn\onecolumn\@tempswatrue\else\@tempswafalse\fi
  \null\vfil
  \markboth{}{}
  {\centering
   \interlinepenalty \@M
   \normalfont
   \MakeUppercase \appendixpagename\par}
  \if@dotoc@pp
    \addappheadtotoc
  \fi
  \vfil\newpage
  \if@twoside
    \if@openright
      \null
      \thispagestyle{empty}
      \newpage
    \fi
  \fi
  \if@tempswa
    \twocolumn
  \fi
}
\makeatother

\definecolor{navycol}{RGB}{100,150,160}
   \definecolor{pinkcol}{RGB}{242,55,55}
   \definecolor{greencol}{RGB}{50,205,50}

   \definecolor{bluecol}{RGB}{30,144,255}

\usepackage{titlesec}

\titleformat*{\section}{\large\bfseries}
\titleformat*{\subsection}{\normalsize\bfseries}
\titleformat*{\subsubsection}{\normalsize\bfseries}
\titleformat*{\paragraph}{\large\bfseries}
\titleformat*{\subparagraph}{\large\bfseries}
\titlespacing{\author}{-5pt}{-5pt}{-5pt}[-5pt]

\makeatletter
\renewcommand\subsubsection{\@startsection{subsubsection}{3}{\z@}
                                     {-3.25ex\@plus -1ex \@minus -.2ex}
                                     {-1.5ex \@plus -.2ex}
                                     {\normalfont\normalsize\bfseries}}
\renewcommand\subsection{\@startsection{subsection}{3}{\z@}
                                     {-3.25ex\@plus -1ex \@minus -.2ex}
                                     {-1.5ex \@plus -.2ex}
                                     {\normalfont\normalsize\bfseries}}                                     
\makeatother

\DeclareFontFamily{U}{solomos}{}
\DeclareErrorFont{U}{solomos}{m}{n}{10}
\DeclareFontShape{U}{solomos}{m}{n}{
  <-> s*[1.1]  gsolomos8r
}{}

   \usepackage{stmaryrd}

\usepackage{tikz}
\usetikzlibrary{arrows.meta}

\usepackage{indentfirst}

\usepackage{tocloft}
\let \savenumberline \numberline
\def \numberline#1{\savenumberline{#1.}}

\usepackage{etoolbox}
\patchcmd{\tableofcontents}{\@starttoc}{\vspace{-0.3cm}\@starttoc}{}{}

\usepackage{accents}
\newcommand\thickbar[1]{\accentset{\rule{.7em}{.9pt}}{#1}}
\newcommand\smallthickbar[1]{\accentset{\rule{.5em}{.7pt}}{#1}}
\usepackage{hyperref}
\hypersetup{
colorlinks=true,
linkcolor=MidnightBlue,
citecolor=violet,
filecolor=purple,
urlcolor=cyan,
breaklinks=true
}

\newcounter{Chapcounter}

\newcommand{\chapter}[1] 
{ {\centering          
  \addtocounter{Chapcounter}{1} \Large \underline{\sffamily \texorpdfstring{\textbf{  Chapter \theChapcounter: ~#1}}{Lg}} }   
  \addcontentsline{toc}{section}{ \color{MidnightBlue} \texorpdfstring{Chapter ~}{Lg}\theChapcounter.\texorpdfstring{~~}{Lg} #1 }    
}

\begin{document}

\title{\vspace{-1.0cm} \textbf{An Excursion \textit{into} Grassmannian Spectra}
\vspace{0.5cm}
}

\author{
 Dmitri Bykov$^{\,a,\,b,\,c,\,d,\,}$\footnote{Emails:
 bykov@mi-ras.ru, dmitri.v.bykov@gmail.com} \quad  Viacheslav Krivorol$^{\,b,\,c,\,}$\footnote{Email:
 v.a.krivorol@gmail.com}
\\  \vspace{0cm}  \\
{\small $a)$ 
\emph{Steklov
Mathematical Institute of Russian Academy of Sciences,}} \\{\small \emph{Gubkina str. 8, 119333 Moscow, Russia} }\\[4pt]
{\small $b)$ 
\emph{Institute for Theoretical and Mathematical Physics,}} \\{\small \emph{Lomonosov Moscow State University, 119991 Moscow, Russia}}\\[4pt]
{\small $c)$ \emph{HSE University, 6 Usacheva str., Moscow 119048, Russia}}\\[4pt]
{\small $d)$ \emph{Beijing Institute of Mathematical Sciences and Applications (BIMSA),}} \\{\small \emph{Huairou District, Beijing
101408, China}}
}

\date{}

{\let\newpage\relax\maketitle}

\maketitle

\vspace{0cm}
\textbf{Abstract.} 
We revisit the classical spectral problem for the Laplace\,--\,Beltrami and Laplace\,--\,Bochner operators on complex Grassmannians. After reviewing the standard approach based on non-commutative harmonic analysis, we develop two complementary methods originating from the quantization of the corresponding one-dimensional sigma model, both with and without a magnetic monopole field. The first is based on $\mathbf{SU}(n)$-equivariant first-order GLSM formulation and gives explicit eigenfunctions in terms of projectors and Pl\"ucker coordinates. The second uses a spin-chain realization of free particles on manifolds and describes the space of square-integrable functions, or sections in the magnetic case, through limiting tensor products of long rectangular $\mathbf{SU}(n)$ representations. Together, these approaches provide a more explicit picture of the spectral geometry of Grassmannians and clarify the relation between harmonics, Young diagrams, magnetic bundles, and Laplacian spectra.

\newpage
\tableofcontents

\newpage

\section{Introduction} 
Laplace-type operators $-\triangle$, namely the Laplace\,--\,Beltrami and Laplace\,--\,Bochner operators, are among the most fundamental elliptic differential operators in Riemannian geometry. On closed compact manifolds, they have discrete non-negative spectra with finite-dimensional eigenspaces, smooth eigenfunctions, and a well-defined heat kernel. Their spectral data encode nontrivial information about the underlying geometry: the Weyl asymptotic formula reflects the dimension and volume of the manifold, while the coefficients of the heat-kernel expansion contain curvature invariants. The study of such relations between geometry and spectral data constitutes the subject of \textit{spectral geometry} \cite{Kac:1966xd,Chavel1984Eigenvalues,Berard1986SpectralGeometry,Rosenberg1997Laplacian}.

A fundamental difficulty of spectral geometry is that, on a generic Riemannian manifold, the spectral problem for the Laplacian is a second-order partial differential equation which usually cannot be solved explicitly. There is, however, an important class of manifolds for which the presence of a large symmetry group makes the problem considerably more tractable, namely homogeneous spaces $\mathbf{G}/\mathbf{H}$ equipped with a $\mathbf{G}$-invariant metric \cite{KobayashiNomizu1996FoundationsVol2}. In this case, the space $L^2\big(\mathbf{G}/\mathbf{H}\big)$, or the corresponding space of square-integrable sections in the magnetic case, carries a representation of $\mathbf{G}$ and can be decomposed into irreducible components. Since the Laplace\,--\,Beltrami and Laplace\,--\,Bochner operators commute with the $\mathbf{G}$-action, their spectral problem can be organized in terms of this representation-theoretic decomposition. Thus, a substantial part of the original differential problem can be reduced to the harmonic analysis and representation theory of $\mathbf{G}$ and $\mathbf{H}$; see \cite{Hel1,Isaev:2018xcg} and the exposition in Section~\ref{HarmonicAnalysisReview}.

However, the case of a general homogeneous space can still be rather complicated. In particular, the irreducible representations appearing in the decomposition of $L^2\big(\mathbf{G}/\mathbf{H}\big)$ may occur with nontrivial multiplicities\footnote{However, substantial progress can still be made for specific classes of manifolds; see, for instance, \cite{Bykov:2024tvb,Bykov:2025lcb,Kuzovchikov:2025mwc} for examples involving flag manifolds with parametric families of invariant metrics.}. Although the Laplace operator acts as a scalar on each irreducible $\mathbf{G}$-representation, it may act nontrivially on the corresponding multiplicity space, so an additional diagonalization is generally required. For compact $\mathbf{G}$, this complication can be avoided for a special choice of $\mathbf{G}$-invariant metric, namely a \textit{normal homogeneous metric}, sometimes also referred to as a \textit{Casimir metric}. For such a metric, the Laplace operator is directly related to the quadratic Casimir operator $\mathcal{C}_2$ of $\mathbf{G}$, schematically 
\begin{align}\label{laplacecas}
    -\triangle=(\mathrm{const})\,\mathcal{C}_2+\mathrm{const}\,.
\end{align}
Since the Casimir acts by the same scalar on all copies of a given irreducible representation, the Laplacian is then automatically diagonal in the multiplicity space as well.

Thus, for a normal homogeneous metric the differential part of the problem essentially trivializes, and one is left with the representation-theoretic problem of determining which irreducible representations, and with which multiplicities, enter the decomposition of the space on which the Laplacian acts. This problem may still be  nontrivial. A further simplification occurs when all multiplicities are either $0$ or $1$, so that one only needs to determine whether a given irreducible representation appears in the decomposition or not. An important class of homogeneous spaces with this property is provided by \textit{symmetric spaces of compact type}. For (irreducible) compact symmetric spaces, the invariant metric is moreover unique up to an overall scale and coincides, with an appropriate normalization, with the normal homogeneous (Casimir) metric \cite{Hel1,Hel2}.

The final simplification is to consider the most rigid and well-controlled class of homogeneous spaces relevant to us, namely \textit{compact Hermitian symmetric spaces}, where complex and algebraic geometry enter the picture. These are compact symmetric spaces equipped with a $\mathbf{G}$-invariant complex structure $J$ compatible with the normal metric $g$. 
Moreover, compact Hermitian symmetric spaces are naturally \textit{K\"ahler}.

\begin{figure}[t]
\centering

\resizebox{0.97\textwidth}{!}{%
\begin{tikzpicture}[
    >={Latex[length=2.5mm]},
    title/.style={
        font=\Large,
        align=center
    },
    layer/.style={
        font=\large,
        align=center
    },
    sub/.style={
        font=\small\itshape,
        text=black!55,
        align=center
    }
]

\definecolor{outerfill}{RGB}{247,245,252}
\definecolor{outerline}{RGB}{145,125,185}

\definecolor{homfill}{RGB}{242,247,252}
\definecolor{homline}{RGB}{105,145,195}

\definecolor{casfill}{RGB}{244,250,244}
\definecolor{casline}{RGB}{100,160,105}

\definecolor{symfill}{RGB}{253,249,235}
\definecolor{symline}{RGB}{200,150,55}

\definecolor{hermfill}{RGB}{253,244,240}
\definecolor{hermline}{RGB}{210,105,70}

\filldraw[
    fill=outerfill,
    draw=outerline,
    line width=0.7pt,
    rounded corners=10pt
]
(0,0) rectangle (14.5,10.2);

\node[title, text width=10cm] at (7.25,9.60)
{Smooth manifolds};

\filldraw[
    fill=homfill,
    draw=homline,
    line width=0.7pt,
    rounded corners=10pt
]
(0.45,0.45) rectangle (14.05,8.70);

\node[layer, text width=10.5cm] at (7.25,8.15)
{Homogeneous spaces $\mathbf{G}/\mathbf{H}$\\[-0.5mm]
with $\mathbf{G}$-invariant metrics}; 

\node[sub, text width=9cm] at (7.25,7.28)
{harmonic analysis / representation theory};

\filldraw[
    fill=casfill,
    draw=casline,
    line width=0.7pt,
    rounded corners=10pt
]
(0.90,0.90) rectangle (13.60,7.00);

\node[layer, text width=10cm] at (7.25,6.30)
{Compact homogeneous spaces with\\[-0.5mm]
normal metric};

\node[sub, text width=9cm] at (7.25,5.48)
{Laplacian determined by the quadratic Casimir};

\filldraw[
    fill=symfill,
    draw=symline,
    line width=0.7pt,
    rounded corners=10pt
]
(1.35,1.35) rectangle (13.15,5.20);

\node[layer, text width=9cm] at (7.25,4.58)
{Symmetric spaces of compact type};

\node[sub, text width=8cm] at (7.25,4.03)
{spherical harmonic analysis};

\filldraw[
    fill=hermfill,
    draw=hermline,
    line width=0.7pt,
    rounded corners=10pt
]
(2.45,1.75) rectangle (12.05,3.78);

\node[layer, text width=8cm] at (7.25,3.08)
{Hermitian symmetric spaces\\[-0.5mm]
of compact type};

\node[sub, text width=7cm] at (7.25,2.18)
{complex, algebraic and K\"ahler geometry};

\draw[
    ->,
    line width=0.8pt,
    black!65
]
(-1.15,8.25) -- (-1.15,1.80);

\node[
    anchor=east,
    align=right,
    text=black!65,
    font=\small,
    text width=3.3cm
] at (-1.55,8.25)
{more general,\\
harder spectral\\
problems};

\node[
    anchor=east,
    align=right,
    text=black!65,
    font=\small,
    text width=3.6cm
] at (-1.55,1.80)
{more rigid,\\
more explicit,\\
\mbox{better controlled}};

\end{tikzpicture}%
}

\caption{A hierarchy of geometric settings for spectral problems.}
\label{GeometricHierarchy}

\end{figure}
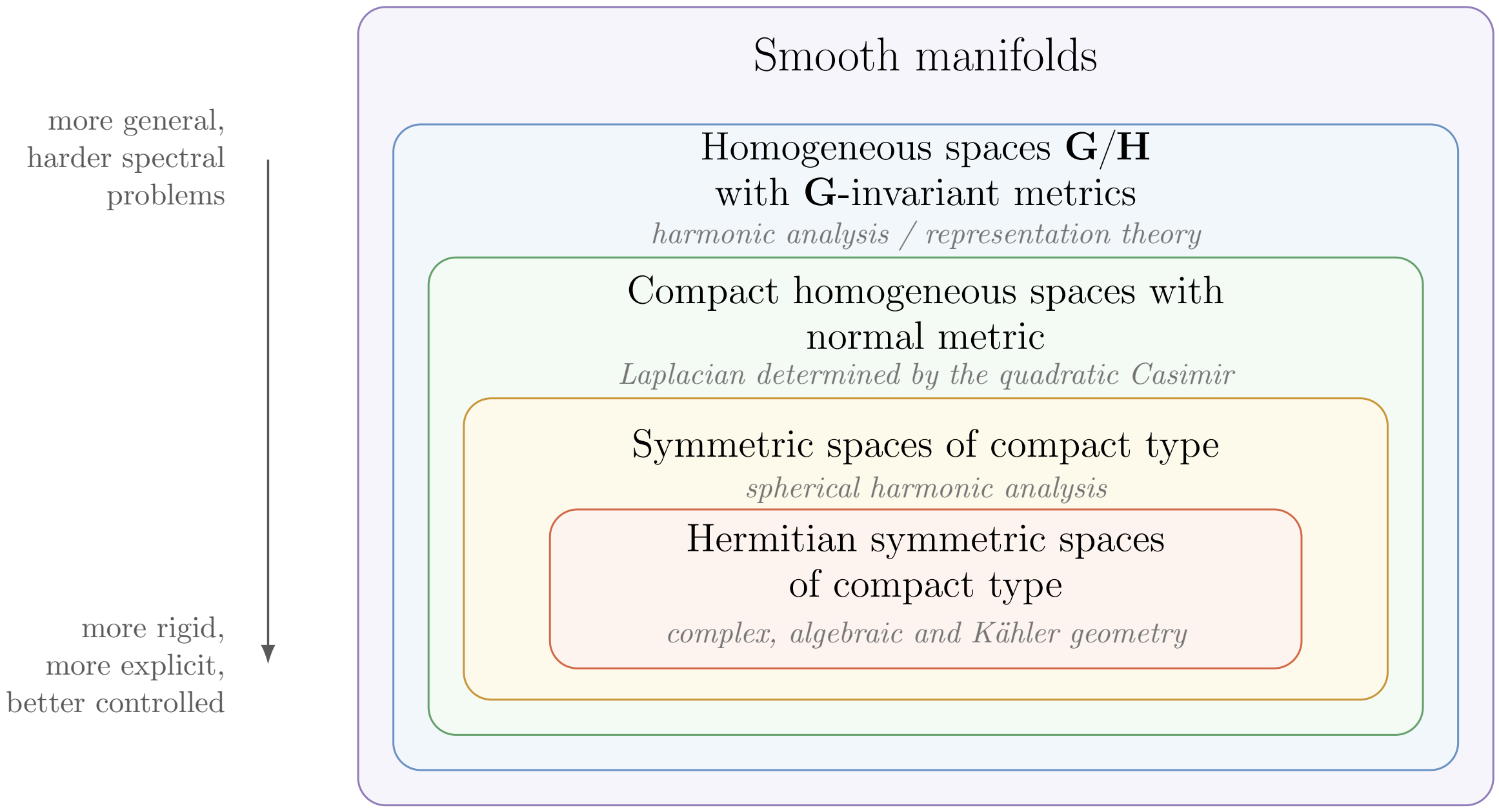

In this work, we focus on the most classical and well-known family of Hermitian symmetric spaces, namely the complex Grassmannians $\mathsf{Gr}(m,n)$, which parametrize $m$-dimensional linear subspaces of a fixed vector space $\mathbb{C}^n$ \cite{GriffithsHarris1978,harris}. Our goal is to solve the corresponding spectral problem in as explicit a form as possible\footnote{See also \cite{BenHalima} for a study of the Grassmannian spectrum, including the magnetic case.}.

Why, then, should one reexamine such a classical problem? There are several reasons. The first is to bridge the gap between the abstract representation-theoretic solution and an explicit description of Grassmannian harmonics. The spectral problem on symmetric spaces is, of course, understood in great generality through the work of Helgason \cite{Hel1,Hel2,Helgason1985FourierTransform}. In this framework, one classifies the highest weights of the irreducible representations entering the decomposition of $L^2\big(\mathbf{G}/\mathbf{H}\big)$, while the corresponding eigenvalues are determined by the quadratic Casimir, as in (\ref{laplacecas}), for instance through the Freudenthal formula \cite[Theorem 2.6]{Knapp1997StructureTheory}.  The eigenfunctions are realized as matrix coefficients of the group elements in the corresponding representations \cite{Isaev:2018xcg} (spherical functions), and the line-bundle case obtained by twisting the equivariance condition by a character. Although this gives a complete and conceptually powerful solution, it remains rather abstract from the geometric point of view: the Grassmannian structure is not manifest, and explicit expressions for the eigenfunctions in natural coordinates are not immediate. One of the main aims of this work is therefore to complement this abstract picture by explicit constructions in which both the geometry of the Grassmannian and the corresponding eigenfunctions become directly visible. In particular, we obtain particularly simple expressions for Grassmannian harmonics in terms of projectors and Pl\"ucker coordinates. Our description is complementary to the known special-function realizations of spherical functions on Grassmannians \cite{BerezinKarpelevich1958Zonal,Hoogenboom1982SphericalFunctions,RemlingRosler2015Convolution}, see \cite{AlHashami2020BerezinKarpelevich} for the magnetic case. The relation between these viewpoints is analogous to the familiar situation on $\mathbb{S}^2$, where spherical harmonics can be written in angular coordinates in terms of associated Legendre functions or, alternatively, realized in a more algebraic and geometric way as restrictions of harmonic homogeneous polynomials on~$\mathbb{R}^3$ \cite{Hall2013QuantumTheory}.

The second reason is that, by solving this spectral problem, we can test a number of techniques and ideas originating in physics, in particular in one-dimensional sigma models and their two-dimensional counterparts \cite{Polyakov:1987hqn}. Intuitively, a one-dimensional sigma model with Riemannian target space $\mathcal{M}$ describes a free (non-relativistic) particle moving on $\mathcal{M}$. Its relation to the spectral problem on $\mathcal{M}$ is very natural: upon quantization, the sigma model Hamiltonian becomes the Laplace\,--\,Beltrami operator (cf.~\cite{Wagner:2021luc}). One may also include a magnetic monopole field, in which case the wavefunctions become sections of a line bundle \cite{Frankel2012GeometryPhysics} and the corresponding Hamiltonian is given by the Laplace\,--\,Bochner operator. The monopole charge is then quantized by the Dirac condition \cite{Alvarez:1984es} and is identified with the first Chern class of the corresponding line bundle \cite{Nakahara2003GeometryTopologyPhysics}.
Point-particle quantum mechanics on curved manifolds, especially in the presence of background magnetic fields, also appears directly in a number of physical contexts. Examples include the integer quantum Hall effect \cite{Karabali:2002im,Wang:2021tic,Klevtsov:2016bos,Balli:2014pqa}, fuzzy geometries \cite{Madore:1991bw,Zhu:2022gjc,He:2026ong,Hasebe:2022pub,Hasebe:2026meg,Dedushenko:2024nwi}, Kaluza\,--\,Klein reduction \cite{Lust1986Compactification}, and many others.

For special classes of target spaces, the corresponding sigma models may admit alternative formulations, which in turn lead to alternative formulations of the spectral problem and can make it more tractable in particular situations. We consider two types of such formulations, both of which are very natural for Grassmannian target spaces.
One of these methods is the \textit{first-order GLSM} (gauged linear sigma model), also known as the \textit{Gross\,--\,Neveu formulation} \cite{Bykov:2021}. From the spectral point of view, its main advantage is that it allows one to formulate the problem on $\mathsf{Gr}(m,n)$ in terms of an analogue of homogeneous coordinates on~$\mathbb{CP}^n$. More precisely, the spectral problem is lifted to the principal $\mathbf{GL}(m,\mathbb{C})$ Stiefel bundle $\mathsf{St}(m,n)\rightarrow\mathsf{Gr}(m,n)$, where $\mathsf{St}(m,n)$ denotes the space of $m$-frames in $\mathbb{C}^n$. In more plain words, the latter is the space of $m\times n$ matrices of rank $m$. This formulation also preserves the $\mathbf{SU}(n)$-equivariance of the problem, which is obscured in a particular Grassmannian chart. We discuss this method in Section~\ref{SectionGLSM}; see also \cite{Krivorol:2025tqj} for a short introduction. 

The second method is the \textit{spin-chain formulation} of point particles. Historically, it originates from Haldane's observation \cite{HALDANE1983464} that the low-energy effective theory of an antiferromagnetic $\mathbf{SU}(2)$ spin chain in the large-spin limit is the $(1+1)$-dimensional sigma model on the sphere $\mathbb{S}^2$, supplemented by a topological $\theta$-term. This relation between spin chains and sigma models was later interpreted geometrically in \cite{Bykov:2011ai,Bykov:2012am} and has recently been developed in a more rigorous form in the one-dimensional setting \cite{Bykov:2024tvb,Bykov:2025lcb,Bykov:2024wft,BykovKrivorol2026HolomorphicQuantizations}; see also \cite{Krivorol:2026egh} for an introduction.
An interesting consequence of this approach is that, in certain cases, the spaces of square-integrable functions $L^2(\mathbf{G}/\mathbf{H})$ or magnetic sections $L_\chi^2(\mathbf{G}/\mathbf{H})$ over a homogeneous space $\mathbf{G}/\mathbf{H}$ can be realized, in an appropriate large-spin limit, in terms of a \textit{finite} number of \textit{tensor products} of $\mathbf{G}$-irreducible representations, which are precisely the Hilbert spaces of finite spin chains. This provides an interesting alternative to the \textit{Peter\,--\,Weyl theorem}, which describes these spaces as \textit{infinite direct sums} over the admissible $\mathbf{G}$-irreducible representations. As we will see for Grassmannians in Section~\ref{LargeSpinSection}, this viewpoint can significantly simplify the organization of the representation-theoretic aspects of the spectral problem.

The paper is organized as follows. In Section~\ref{HarmonicAnalysisReview}, we review some basic aspects of harmonic analysis on homogeneous spaces and their application to Grassmannians. In Section~\ref{SectionGLSM}, we solve the Grassmannian spectral problem by a completely different method based on the first-order GLSM formulation. There, we not only derive the eigenvalues in a relatively straightforward way, but also construct explicit Grassmannian harmonics, including the magnetic case, in ``homogeneous'', or ``Stiefel'', coordinates. Finally, in Section~\ref{LargeSpinSection}, we derive the eigenspace decomposition once again by representing it as the large-spin limit of a tensor product of two ``infinitely long'' rectangular $\mathbf{SU}(n)$ representations.

\section{Aspects of harmonic analysis}\label{HarmonicAnalysisReview}
In this section, for the sake of self-consistency, we review some textbook material needed to understand the group- and representation-theoretic approach to the spectral problem, together with several classical results that form the basis of noncommutative harmonic analysis. Our exposition includes the realization of functions and sections of line bundles on homogeneous spaces, the relation between Laplace operators and the quadratic Casimir, and the solution of the spectral problem in terms of spherical weights. We illustrate these techniques explicitly for Grassmannians.

\subsection{Functions and line bundles over homogeneous spaces.}\label{FunOnHomSubsection}
We begin by recalling how functions and sections of line bundles over a homogeneous space\footnote{Here and hereafter we assume that both groups are compact and $\mathbf{G}$ is semisimple.} $\mathbf{G}/\mathbf{H}$ can be represented in terms of functions on $\mathbf{G}$. Since the points of $\mathbf{G}/\mathbf{H}$ are right cosets, a point is represented by an element $g\in\mathbf{G}$ modulo the equivalence relation $g\sim gh$, with $h\in\mathbf{H}$. Accordingly, functions on $\mathbf{G}/\mathbf{H}$ can be identified with right $\mathbf{H}$-invariant functions on $\mathbf{G}$ under the right action,
\begin{equation}
L^2\big(\mathbf{G}/\mathbf{H}\big)
\simeq
L^2(\mathbf{G})^{\mathbf{H}}
\simeq
\Big\{f(g):\mathbf{G}\rightarrow\mathbb{C}~\Big|~f(gh)=f(g),\quad g\in\mathbf{G}\,,\quad h\in\mathbf{H}\Big\}\,.
\end{equation}

The $\mathbf{G}$-equivariant line bundles can be described in a similar way, with a slight modification. It is useful to regard them as line bundles associated, through a group character, with the principal bundle $\mathbf{G}\rightarrow\mathbf{G}/\mathbf{H}$. Namely, for a fixed character $\chi:\mathbf{H}\rightarrow\mathbf{U}(1)$, the corresponding space of square-integrable sections can be represented by equivariant functions on $\mathbf{G}$,
\begin{equation}
L_\chi^2\big(\mathbf{G}/\mathbf{H}\big):=\mathrm{Ind}^{\mathbf{G}}_{\mathbf{H}}\chi
\simeq
\Big\{f(g)~\Big|~f(gh)=\chi^{-1}(h)f(g),\quad g\in\mathbf{G}\,,\quad h\in\mathbf{H}\Big\}\,.
\end{equation}

The spaces $L^2\big(\mathbf{G}/\mathbf{H}\big)$ and $L_\chi^2\big(\mathbf{G}/\mathbf{H}\big)$ carry natural representations of $\mathbf{G}$ through the left action and can therefore be decomposed into irreducible $\mathbf{G}$-representations. This decomposition follows directly from the Peter\,--\,Weyl theorem \cite[Section 35.4]{Etingof:2022gcn}. Namely, $L^2(\mathbf{G})$, regarded as a representation of $\mathbf{G}\times\mathbf{G}$ through the commuting left and right actions, decomposes as
\begin{equation}
L^2(\mathbf{G})
\simeq
\bigoplus\limits_{\rho}
\mathsf{V}_\rho\otimes\mathsf{V}_\rho^\ast\,,
\end{equation}
where $\rho$ runs over the inequivalent irreducible representations of $\mathbf{G}$, $\mathsf{V}_\rho$ carries the left action, and $\mathsf{V}_\rho^\ast$ carries the corresponding right action.
To pass to the homogeneous space, we restrict the right factor to the subspace transforming according to the character $\chi$. More precisely, let
\begin{equation}
\big(\mathsf{V}_\rho^\ast\big)_\chi
=
\Big\{
v\in\mathsf{V}_\rho^\ast
\quad\Big|\quad
\rho^\ast(h)v=\chi^{-1}(h)v\,,\quad h\in\mathbf{H}
\Big\}\,.
\end{equation}
Thus, 
\begin{equation}
L_\chi^2\big(\mathbf{G}/\mathbf{H}\big)
\simeq
\bigoplus\limits_{\rho}
\mathsf{V}_\rho\otimes
\big(\mathsf{V}_\rho^\ast\big)_\chi\,.
\end{equation}
For the trivial character $\chi(h)=1$, the space
$\big(\mathsf{V}_\rho^\ast\big)_\chi$ reduces to the space of
$\mathbf{H}$-invariant vectors $\big(\mathsf{V}_\rho^\ast\big)^{\mathbf{H}}$,
and therefore
\begin{equation}\label{PeterWeyl}
L^2\big(\mathbf{G}/\mathbf{H}\big)
\simeq
\bigoplus\limits_{\rho}
\mathsf{V}_\rho\otimes
\big(\mathsf{V}_\rho^\ast\big)^{\mathbf{H}}\,.
\end{equation}

\subsection{Casimir operator.}
Now let us describe the differential realization of the quadratic Casimir operator $\mathcal{C}_2$ of $\mathbf{G}$ on $L_\chi^2\big(\mathbf{G}/\mathbf{H}\big)$. For an element $A\in\mathfrak{g}=\operatorname{Lie}(\mathbf{G})$, consider the differential operator generated by infinitesimal right translations\footnote{Here and below we assume that the functions involved are sufficiently differentiable.},
\begin{equation}
R_A f(g) := \frac{\dif}{\dif t} f\Big(g e^{tA}\Big)
\bigg|_{t=0}\,.
\end{equation}
If $\{X_a\}$ is an orthonormal basis of $\mathfrak{g}$ with respect to the Killing metric, then the quadratic Casimir $\mathcal{C}_2(\mathbf{G}) = X^a X_a$ is realized on $L^2(\mathbf{G})$ as the second-order differential operator
\begin{equation}\label{GroupCasimirAction}
\mathcal{C}_2(\mathbf{G}) f(g)=-\sum\limits_{a=1}^{\dim\mathfrak{g}}R_{X_a}^2 f(g)\,,\qquad f(g)\in L^2(\mathbf{G})\,.
\end{equation}
For the coset space $\mathbf{G}/\mathbf{H}$, we have the orthogonal decomposition $\mathfrak{g}
=
\mathfrak{h}\oplus\mathfrak{m}$,
where $\mathfrak{h}=\operatorname{Lie}(\mathbf{H})$ and $\mathfrak{m}=\mathfrak{h}^{\perp}$ is the orthogonal complement of $\mathfrak{h}$ in $\mathfrak{g}$ with respect to the Killing metric. Choosing orthonormal bases $\{H_a\}$ of $\mathfrak{h}$ and $\{M_a\}$ of $\mathfrak{m}$, the quadratic Casimir of $\mathbf{G}$ decomposes as
\begin{equation}\label{FullCasimirOnHomogeneousSpace}
\mathcal{C}_2(\mathbf{G})
=
-\sum\limits_{a=1}^{\dim\mathfrak{m}}R_{M_a}^2
-
\sum\limits_{a=1}^{\dim\mathfrak{h}}R_{H_a}^2\,.
\end{equation}
Upon restriction from $L^2(\mathbf{G})$ to $L^2(\mathbf{G}/\mathbf{H})$, the $\mathfrak{h}$-part vanishes, since functions on $\mathbf{G}/\mathbf{H}$ are represented by right $\mathbf{H}$-invariant functions on $\mathbf{G}$ and therefore satisfy $R_Hf(g)=0$ for every $H\in\mathfrak{h}$. Thus,
\begin{equation}
\mathcal{C}_2(\mathbf{G}) f(g)
=
-\sum\limits_{a=1}^{\dim\mathfrak{m}}R_{M_a}^2 f(g)\,,\qquad f(g)\in L^2\big(\mathbf{G}/\mathbf{H}\big)\,.
\end{equation}

The magnetic case $L_\chi^2\big(\mathbf{G}/\mathbf{H}\big)$ is slightly more involved. In this case, the second term in \eqref{FullCasimirOnHomogeneousSpace} does not vanish upon restriction to $L_\chi^2\big(\mathbf{G}/\mathbf{H}\big)$, but instead reduces to a constant. Indeed, using the equivariance condition $f(gh)=\chi(h)^{-1}f(g)$, we find
\begin{equation}\label{MagneticConstantBochner}
\sum\limits_{a=1}^{\dim\mathfrak{h}}R_{H_a}^2 f(g)
=
\sum\limits_{a=1}^{\dim\mathfrak{h}}
\Big(\dif\chi\big(H_a\big)\Big)^2 f(g)\,,
\quad
f\in L_\chi^2\big(\mathbf{G}/\mathbf{H}\big)\,,
\quad
\dif\chi(H)
:=
\frac{\dif}{\dif t}
\chi\Big(e^{tH}\Big)\bigg|_{t=0}\,.
\end{equation}
Thus, the contribution of the $\mathfrak{h}$-part of the Casimir is determined by the squared Killing norm on $\mathfrak{g}$ of the character differential $\dif\chi:\mathfrak{h}\rightarrow\mathfrak{u}(1)\simeq i\mathbb{R}$, i.e.
\begin{equation}
\lVert\dif\chi\rVert^2
:=
-\sum\limits_{a=1}^{\dim\mathfrak{h}}
\Big(\dif\chi\big(H_a\big)\Big)^2\,.
\end{equation}
This quantity may equivalently be interpreted as the eigenvalue $\mathcal{C}_2(\mathbf{H};\chi)$ of the quadratic Casimir of $\mathfrak{h}$, defined using the restriction to $\mathfrak{h}$ of the invariant inner product on $\mathfrak{g}$, in the one-dimensional representation defined by the character $\chi$.

Let us now compute this constant for the Grassmannians $\mathsf{Gr}(m,n)$. In this case,
\begin{equation}
\mathfrak{g}\simeq\mathfrak{su}(n)\,,\qquad
\mathfrak{h}\simeq\mathfrak{s}\big(\mathfrak{u}(m)\oplus\mathfrak{u}(n-m)\big)
\simeq
\mathfrak{su}(m)\oplus\mathfrak{su}(n-m)\oplus\mathfrak{u}(1)\,.
\end{equation}
The key observation is that, in computing the norm of $\dif\chi$, only the $\mathfrak{u}(1)$ component of $\mathfrak{h}$ contributes, since $\dif\chi$ vanishes on the semisimple components\footnote{Indeed, for a semisimple Lie algebra $\mathfrak{g}$, the orthogonal complement $[\mathfrak{g},\mathfrak{g}]^\perp$ with respect to the Killing form coincides with the center of $\mathfrak{g}$ by $\mathrm{ad}_{\mathfrak{g}}$-invariance. Since the center is an abelian ideal, it is trivial for a semisimple Lie algebra, and hence $\mathfrak{g}=[\mathfrak{g},\mathfrak{g}]$. On the other hand, $\dif\chi$ vanishes on commutators since it is a Lie algebra homomorphism to the abelian algebra $\mathfrak{u}(1)$.}. This $\mathfrak{u}(1)$ component is generated by
\begin{equation}
Z
:=
i\begin{pmatrix}
\frac{\mathds{1}_m}{m} & 0\\
0 & -\frac{\mathds{1}_{n-m}}{n-m}
\end{pmatrix}\in\mathfrak{h}\,.
\end{equation}
We normalize the invariant inner product on $\mathfrak{g}=\mathfrak{su}_n$ as $\langle X,Y\rangle=-2\mathrm{Tr}(XY)$, where $X$ and~$Y$ are $n\times n$ matrices. Then
\begin{equation}
\lVert Z\rVert^2
=
\frac{2}{m}+\frac{2}{n-m}
=
\frac{2n}{m(n-m)}\,,
\end{equation}
and hence the normalized $\mathfrak{u}(1)$ generator is
\begin{equation}
H_{\mathfrak{u}(1)}
=
\sqrt{\frac{m(n-m)}{2n}}\,Z\,.
\end{equation}
The character with monopole charge $\mathfrak{q}\in\mathbb{Z}$ is characterized by
\begin{equation}
\chi_{\mathfrak{q}}\big(e^{tZ}\big)
=
e^{i\mathfrak{q}t}\,,
\qquad
\dif\chi_{\mathfrak{q}}(Z)
=
i\mathfrak{q}\,.
\end{equation}
Therefore,
\begin{equation}\label{BochnerGrassmannDifference}
\lVert\dif\chi_{\mathfrak{q}}\rVert^2
=
-\Big(\dif\chi\big(H_{\mathfrak{u}(1)}\big)\Big)^2
=
\frac{\mathfrak{q}^2m(n-m)}{2n}\,.
\end{equation}
Thus, on $L_{\chi}^2\big(\mathsf{Gr}(m,n)\big)$, the quadratic Casimir takes the form
\begin{equation}
\mathcal{C}_2(\mathbf{G})
=
-\sum\limits_{a=1}^{\dim\mathfrak{m}}R_{M_a}^2
+
\frac{\mathfrak{q}^2m(n-m)}{2n}\,.
\end{equation}

\subsection{Normal metric.}
In this section, we clarify the relation between the Laplace and Casimir operators on homogeneous spaces equipped with a normal metric. For pedagogical reasons, we start with the case of bi-invariant metric on $\mathbf{G}$.

Consider first the group $\mathbf{G}$. Let $\{X_a\}$ be a basis of $\mathfrak{g}=\mathrm{T}_e\mathbf{G}$. The vector field $R_{X_a}$ is obtained by pushing forward $X_a$ from the identity to a point $g\in\mathbf{G}$ by the left translation $L_g$, namely $R_{X_a}(g)=\dif L_g(X_a):=(L_g)_\ast X_a$. Since $L_g$ is a diffeomorphism, the vectors $\{R_{X_a}(g)\}$ form a basis of $\mathrm{T}_g\mathbf{G}$ at every point $g\in\mathbf{G}$. Thus, denoting $E_a:=R_{X_a}$, we obtain a global frame $\{E_a\}$ on $\mathbf{G}$. We can also introduce the dual coframe $\{e^a\}$, consisting of one-forms defined by $e^a(E_b)=\delta^a_b$.

It is also useful to introduce an object inverse, in a sense, to the above construction. While $(L_g)_\ast$ maps vectors from $\mathfrak{g}=\mathrm{T}_e\mathbf{G}$ to $\mathrm{T}_g\mathbf{G}$, the differential of the inverse left translation maps them in the opposite direction, $(L_{g^{-1}})_\ast:\mathrm{T}_g\mathbf{G}\rightarrow\mathfrak{g}$. Thus, at every point $g\in\mathbf{G}$ we define the $\mathfrak{g}$-valued one-form $\theta$ by $\theta_g:=(L_{g^{-1}})_\ast$. This is the \textit{Maurer\,--\,Cartan one-form} (cf.~\cite{Wise:2006sm}). In physics, people usually prefer the matrix group notation \cite{OConnor:2024lbe} $\theta_g:=g^{-1}\dif g$, because in natural exponential parametrization $g(t) = g(0)e^{t A}$ it gives $g^{-1}(t)\dif g(t) = A\,\dif t$. It is inverse to $R$ in the sense that $\theta_g\big(R_{X_a}(g)\big)=(L_{g^{-1}})_\ast(L_g)_\ast X_a=X_a$. Therefore, we obtain $\theta=e^aX_a$. 

The Maurer\,--\,Cartan form is particularly useful for defining a bi-invariant metric on the group. Let $\langle\bullet,\bullet\rangle$ be an $\mathrm{Ad}(\mathbf{G})$-invariant inner product on $\mathfrak{g}$, for instance a suitably normalized negative Killing form. Since $\theta=g^{-1}\dif g$ is $\mathfrak{g}$-valued, left-invariant, and transforms by the adjoint action under right translations, the metric $G=\langle\theta,\theta\rangle=\langle g^{-1}\dif g,g^{-1}\dif g\rangle$ is invariant under both left and right translations. Equivalently, substituting $\theta=e^aX_a$ for an orthonormal basis $\{X_a\}$ of $\mathfrak{g}$, we obtain $G=\delta_{ab}e^ae^b$, while the inverse metric is $G^{-1}=\delta^{ab}E_aE_b$.

Let us now compute the corresponding Laplace\,--\,Beltrami operator. Introducing local coordinates $x^\mu$ on $\mathbf{G}$, it takes the standard form
\begin{equation}
-\triangle_{\mathbf{G}} f
=
-\frac{1}{\sqrt{\det G}}\,
\partial_\mu\Big(\sqrt{\det G}\,
G^{\mu\nu}\partial_\nu f\Big)\,,
\qquad
f\in L^2(\mathbf{G})\,.
\end{equation}
Writing $E_a=E_a^\mu\partial_\mu$, we have $G^{\mu\nu}=\delta^{ab}E_a^\mu E_b^\nu$, and therefore
\begin{equation}
-\triangle_{\mathbf{G}} f
=
-\sum\limits_{a = 1}^{\mathrm{dim}\mathfrak{g}} E_a^2f
-
\frac{1}{\sqrt{\det G}}\,
\partial_\mu\big(\sqrt{\det G}\,E_a^\mu\big)
E_a^\nu\partial_\nu f\,.
\end{equation}
The first term is precisely the quadratic Casimir action \eqref{GroupCasimirAction}, while the second term vanishes due to the incompressibility condition $\mathrm{div}_G R_{X_a}=0$, which is a consequence of the fact that $E_a=R_{X_a}$ are Killing vector fields. 
Thus, we conclude that
\begin{equation}
-\triangle_{\mathbf{G}} f(g)
=
\mathcal{C}_2(\mathbf{G})f(g)\,,
\qquad
f\in L^2(\mathbf{G})\,.
\end{equation}
This identification also makes the $\mathbf{G}$-invariance of the Laplace operator manifest and explains why representation-theoretic methods are so effective in the spectral problem.

The generalization to the coset space $\mathbf{G}/\mathbf{H}$ is very natural. Using the orthogonal decomposition $\mathfrak{g}=\mathfrak{h}\oplus\mathfrak{m}$, the Maurer\,--\,Cartan form decomposes accordingly as $\theta=\theta_{\mathfrak{h}}+\theta_{\mathfrak{m}}$. The $\mathfrak{m}$-component describes the tangent directions on the coset, and the normal metric on $\mathbf{G}/\mathbf{H}$ is defined by requiring its pullback under the canonical projection $\pi:\mathbf{G}\rightarrow\mathbf{G}/\mathbf{H}$ to be $\langle\theta_{\mathfrak{m}},\theta_{\mathfrak{m}}\rangle$. At the same time, $\theta_{\mathfrak{h}}$ defines the canonical connection \cite{Sharpe1997DifferentialGeometry} on the principal bundle $\mathbf{G}\rightarrow\mathbf{G}/\mathbf{H}$.
Splitting an orthonormal basis $\{X_a\}$ of $\mathfrak{g}$ into bases $\{H_a\}$ of $\mathfrak{h}$ and $\{M_a\}$ of $\mathfrak{m}$, the Laplace\,--\,Beltrami operator for the normal metric, as well as the Laplace\,--\,Bochner operator associated with the canonical connection, is represented by the $\mathfrak{m}$-part of the quadratic Casimir \cite{Hu2008TheLO},
\begin{equation}
-\triangle_{\mathbf{G}/\mathbf{H}}
=
-\sum\limits_{a=1}^{\dim\mathfrak{m}}R_{M_a}^2\,.
\end{equation}
Here $R_{M_a}$ are horizontal vector fields on $\mathbf{G}$ for the principal connection $\theta_{\mathfrak{h}}$. Acting on the $\mathbf{H}$-equivariant functions representing sections of the associated line bundle, they give the components of the induced covariant derivative. Throughout this paper we consider line bundles associated with unitary characters $\chi:\mathbf{H}\rightarrow\mathbf{U}(1)$, although the same formula holds for vector bundles associated with finite-dimensional unitary representations of $\mathbf{H}$. The invariant operator $-\sum_a R_{M_a}^2$ preserves equivariance and represents the Bochner Laplacian on sections, reducing to the Laplace\,--\,Beltrami operator for the trivial character.

Taking into account the contribution \eqref{MagneticConstantBochner} of the $\mathfrak{h}$-part of the Casimir in the magnetic case, we finally obtain \cite[Appendix A.2]{Meng2003}
\begin{equation}\label{laplacecas2}
-\triangle_{\mathbf{G}/\mathbf{H}}^{\textit{Beltr}}
=
\mathcal{C}_2(\mathbf{G})
\quad\text{on}\quad
L^2(\mathbf{G}/\mathbf{H})\,,
\quad
-\triangle_{\mathbf{G}/\mathbf{H}}^{\textit{Bochn}}
=
\mathcal{C}_2(\mathbf{G})-\lVert\dif\chi\rVert^2
\quad\text{on}\quad
L_\chi^2(\mathbf{G}/\mathbf{H})\,.
\end{equation}

\subsection{Spectrum from spherical weights.}
In Section~\ref{FunOnHomSubsection}, we discussed the space of functions $L^2(\mathbf{G}/\mathbf{H})$ and its representation-theoretic structure under the left action of $\mathbf{G}$. In particular, $L^2(\mathbf{G}/\mathbf{H})$ is a reducible representation of $\mathbf{G}$, whose decomposition into irreducible representations $\mathsf{V}_\rho$ is given by the Peter\,--\,Weyl theorem \eqref{PeterWeyl}. If $\dim\big(\mathsf{V}_\rho^\ast\big)^\mathbf{H}\neq0$, the corresponding representation $\mathsf{V}_\rho$ is called $\mathbf{H}$-\textit{spherical}. If the representation $\mathsf{V}_\rho$ is spherical, then the corresponding highest weight is also called spherical.

Once the Peter\,--\,Weyl decomposition is established, one may ask whether it is multiplicity-free, i.e.~whether $\dim\big(\mathsf{V}_\rho^\ast\big)^\mathbf{H}\leq1$. In particular, this is the case for \textit{symmetric spaces} \cite{Helgason1985FourierTransform}. Thus, in this case the decomposition \eqref{PeterWeyl} simplifies to
\begin{equation}
L^2\big(\mathbf{G}/\mathbf{H}\big)\simeq\bigoplus_{\text{spherical~}\rho} \mathsf{V}_\rho\,.
\end{equation}
An explicit parametrization of the spherical weights for all symmetric spaces in terms of fundamental weights can be found in \cite[pages 29 and 30]{GoodmanWallach1996AlgebraicGroup}.

Since the Casimir eigenvalue on $\mathsf{V}_\rho$ is in principle known, in order to find the Laplacian eigenvalues it is enough to describe the spherical representations. They are often parametrized not in terms of Young diagrams but in terms of highest weights, so it is useful to express the Casimir eigenvalues in these terms. They are given by \textit{Freudenthal's formula}~\cite{Knapp1997StructureTheory}
\begin{equation}\label{Freudenthal}
\mathcal{C}_2(\rho) = \langle\rho,\rho+\delta\rangle\,,
\end{equation}
where $\rho$ is the dominant integral highest weight which defines the corresponding $\mathsf{V}_\rho$. Also, $\delta$ is the sum over all positive roots of $\mathbf{G}$ and $\langle\bullet,\bullet\rangle$ is an invariant scalar product on the weight space.

\subsubsection{Perelomov\,--\,Popov formula.}
The quadratic Casimir eigenvalue can alternatively be expressed directly in terms of the Young diagram specifying the representation. For $\mathbf{SU}(n)$, this is given the \textit{Perelomov\,--\,Popov formula} \cite{PerelomovPopov1966CasimirUnSU}. Here we derive it from the Freudenthal formula. Let $R_1\geq\ldots\geq R_n\geq0$ be the row lengths of the Young diagram $\mathbb{Y}$, with $R_k=0$ for missing rows. These lengths give a representative $\sum_{k=1}^n R_k\varepsilon_k$ of the $\mathfrak{su}(n)$ highest weight, where $\varepsilon$ is the $\varepsilon$-functional on the $\mathfrak{u}(n)$ Cartan subalgebra. Subtracting the overall averaging constant, we can write the highest weight as
\begin{equation}
\rho=\sum\limits_{k=1}^n s_k\varepsilon_k\,,
\qquad
s_k=R_k-\frac1n\sum\limits_{j=1}^n R_j\,,
\qquad
\sum\limits_{k=1}^n s_k=0\,.
\end{equation}
Substituting this expression into the Freudenthal formula \eqref{Freudenthal}, with the $\mathfrak{su}(n)$ positive root sum $\delta=\sum_{k=1}^n(n+1-2k)\varepsilon_k$ and a normalization $\langle\varepsilon_i,\varepsilon_j\rangle=\frac12\delta_{ij}$, gives
\begin{equation}
\mathcal{C}_2(\rho)
=
\langle\rho,\rho\rangle+\langle\rho,\delta\rangle
=
\frac12\sum\limits_{k=1}^n s_k^2
+
\frac12\sum\limits_{k=1}^n(n+1-2k)s_k\,.
\end{equation}
The contribution $\frac{n+1}{2}\sum_{k=1}^n s_k$ vanishes, leaving the Perelomov\,--\,Popov formula\footnote{A direct derivation using the Schwinger\,--\,Wigner oscillator realization can also be found in~\cite{AffleckBykov:2021}.}
\begin{equation}
\label{PerelomovPopov}
\mathcal{C}_2(\mathbb{Y})
=
\frac12\sum\limits_{k=1}^n s_k(s_k-2k)\,.
\end{equation}
We will use this formula in  Section \ref{LargeSpinSection} below.

\subsubsection{Spherical functions.}
In the case of a symmetric space, one can also construct the Laplace\,--\,Beltrami eigenfunctions in an invariant representation-theoretic form, although making these expressions explicit in local coordinates may still be challenging. This construction is naturally described in terms of \textit{spherical functions} \cite{Helgason1985FourierTransform,Isaev:2018xcg}. Let $\mathsf{V}_\rho$ be an $\mathbf{H}$-spherical irreducible representation of $\mathbf{G}$. For a symmetric space, the $\mathbf{H}$-invariant subspace is one-dimensional, $\dim\mathsf{V}_\rho^{\mathbf{H}}=1$, so we may choose a normalized invariant vector $v_\rho^{\mathbf{H}}\in\mathsf{V}_\rho^{\mathbf{H}}$. Then, for any $w\in\mathsf{V}_\rho$, the corresponding matrix coefficient
\begin{equation}
\psi_w(g\mathbf{H})
=
\big\langle w,g\cdot v_\rho^{\mathbf{H}}\big\rangle
\end{equation}
defines a function on $\mathbf{G}/\mathbf{H}$, where $\langle\bullet,\bullet\rangle$ is the $\mathbf{G}$-invariant Hermitian product. As $w$ varies over $\mathsf{V}_\rho$, these functions span the copy of $\mathsf{V}_\rho$ appearing in $L^2(\mathbf{G}/\mathbf{H})$ and have the same Laplace\,--\,Beltrami eigenvalue determined by the quadratic Casimir $\mathcal{C}_2(\rho)$. For example, on the sphere $\mathbb{S}^2\simeq\mathbf{SO}(3)/\mathbf{SO}(2)$, this construction gives the standard spherical harmonics $Y_{\ell}^{m}(\theta,\varphi)$.

A distinguished member of this family is obtained by choosing $w=v_\rho^{\mathbf{H}}$. The resulting function
\begin{equation}
\phi_\rho(g)
=
\big\langle v_\rho^{\mathbf{H}},g\cdot v_\rho^{\mathbf{H}}\big\rangle
\end{equation}
is called the \textit{zonal spherical function}. It is invariant under both left and right multiplication by $\mathbf{H}$,
\begin{equation}
\phi_\rho(h_1gh_2)=\phi_\rho(g)\,,\qquad h_1,h_2\in\mathbf{H}\,,
\end{equation}
and is normalized by $\phi_\rho(e)=1$, where $e$ is the unit element. Geometrically, it is the unique normalized eigenfunction in the representation $\mathsf{V}_\rho$ which is invariant under the isotropy group of the base point $e\mathbf{H}$. In this sense, the zonal spherical function contains the radial part of the corresponding eigenspace, while the remaining eigenfunctions are obtained by the action of $\mathbf{G}$. 
For example, on the sphere, the zonal spherical functions are proportional to the spherical harmonics $Y_{\ell}^0(\theta)$, which are independent of the azimuthal angle $\varphi$.

\subsubsection{Calculation of Grassmannian spectrum.}
Let us explicitly demonstrate how, by using methods described above, one can find the Laplace\,--\,Beltrami spectrum in the case\footnote{Here we assume that $n\geq 2m$ because $\mathsf{Gr}(n-m,n)\simeq\mathsf{Gr}(m,n)$.} of the Grassmannians $\mathsf{Gr}(m,n)$. They are the most well-known symmetric spaces, and according to the table in \cite{GoodmanWallach1996AlgebraicGroup} the generators of the corresponding spherical weights are
\begin{equation}
\mu_k:=\omega_k + \omega_{n-k}\,,\qquad k = 1,\ldots,m\,,
\end{equation}
where
\begin{equation}
\omega_k = \varepsilon_1+\ldots+\varepsilon_k - \frac{k}{n}\big(\varepsilon_1+\ldots+\varepsilon_n\big)\,,\qquad \varepsilon_i\big(\mathrm{diag}(x_1,\ldots,x_n)\big) = x_i
\end{equation}
are the fundamental weights of $\mathbf{SU}(n)$, given in terms of the standard $\varepsilon$-functionals on the $\mathfrak{u}(n)$ Cartan subalgebra. Thus,
\begin{align}
\mu_k &= \Bigg(\sum\limits_{i = 1}^k \varepsilon_i - \frac{k}{n}\sum\limits_{i = 1}^n \varepsilon_i\Bigg) + \Bigg(\sum\limits_{i = 1}^{n-k} \varepsilon_i - \frac{n-k}{n}\sum\limits_{i = 1}^n \varepsilon_i\Bigg) =\sum\limits_{i = 1}^k \varepsilon_i+\Bigg(\sum\limits_{i = 1}^{n-k}\varepsilon_i - \sum\limits_{i = 1}^n \varepsilon_i\Bigg)=\nonumber\\
&=\sum\limits_{i = 1}^k\big(\varepsilon_i - \varepsilon_{n-k+i}\big) = \varepsilon_1+\ldots+\varepsilon_{k}-\varepsilon_{n-k+1}-\ldots-\varepsilon_n\,.
\end{align}
This means that every spherical weight $\rho$ for the Grassmannian has the form
\begin{equation}\label{GrassmWeight1}
\rho = \sum\limits_{k=1}^m\lambda_k\mu_k = \sum\limits_{k = 1}^m\lambda_k\sum\limits_{i = 1}^k\big(\varepsilon_i - \varepsilon_{n-k+i}\big)\,,\qquad \lambda_k\in \mathbb{Z}_{\geq 0}\,.
\end{equation}
It is more natural to introduce here a parametrization by $L_k:=\lambda_k+\ldots+\lambda_m$, i.e.
\begin{equation}\label{GrassmWeight2}
\rho = L_1\varepsilon_1 +\ldots+L_m\varepsilon_m - L_m\varepsilon_{n-m+1}-\ldots-L_1\varepsilon_n\,,\qquad L_1\geq L_2\geq\ldots\geq L_m\geq 0\,,
\end{equation}
which coincides with our further result \eqref{HighestWeightSU}, as it should. As we will see later on, $L_k$ are the lengths of the rows of a certain auxiliary Young diagram, see~(\ref{AntiHolomorphicYoungDiagramGr}).

For $\mathbf{SU}(n)$, the positive roots have the form $\alpha_{ij} = \varepsilon_i - \varepsilon_j$ with $i<j$. The sum of all positive roots is
\begin{equation}
\delta = \sum\limits_{1\leq i<j\leq n}\big(\varepsilon_i - \varepsilon_j \big) = \sum\limits_{k=1}^n (n+1-2k)\varepsilon_k\,.
\end{equation}
Note that the coefficients of $\varepsilon_k$ and its partner $\varepsilon_{n-k+1}$ in $\delta$ differ only by a sign. Thus, the quadratic Casimir eigenvalue $\mathcal{C}_2(\rho) = \langle\rho,\rho\rangle + \langle\rho,\delta\rangle$
is given by
\begin{equation}\label{SphericalCasimirGr}
\langle\rho,\rho\rangle = \sum\limits_{k = 1}^m L_k^2\,,\quad \langle\rho,\delta\rangle = \sum\limits_{k = 1}^m L_k (n+1-2k)\,,\quad \mathcal{C}_2(\rho) = \sum\limits_{k=1}^m L_k\big(L_k+n-2k+1\big)\,,
\end{equation}
where we normalize the scalar product as $\langle\varepsilon_i,\varepsilon_j\rangle = \frac{1}{2}\delta_{ij}$. 
Note that the contributions of $\varepsilon_k$ and $\varepsilon_{n-k+1}$ to the Casimir eigenvalue are equal. As a result, we have derived, via~(\ref{laplacecas2}), the spectrum  of the Grassmannian.

\subsubsection{Magnetic spectrum.}
The generalization of this result to the magnetic case is rather simple. The magnetic analogues of \eqref{GrassmWeight1} and \eqref{GrassmWeight2} can be read off from\footnote{Note that the authors of this reference use restricted weights, i.e.~the restrictions of the ordinary weights to a maximal abelian subspace (Cartan subspace) of $\mathfrak{m}$.}
\cite[Theorem~3.1 and Proposition~3.3]{Olafsson}; the underlying general result was proved in \cite{Schlichtkrull1984OneDimensionalKTypes}. Translating their result into our notation, the magnetic spherical weights are given by
\begin{equation}\label{MagnWeight}
\rho_{\mathfrak{q}}
=
\mathfrak{q}\omega_{n-m}
+
\sum\limits_{k=1}^m\lambda_k\mu_k:=\mathfrak{q}\omega_{n-m}
+\rho_0\,,
\end{equation}
where we assume $\mathfrak{q}\geq0$. 
Equivalently, in terms of $L_k=\lambda_k+\ldots+\lambda_m$, this weight can be written for $\mathfrak{su}(n)$ as
\begin{equation}
\rho_{\mathfrak{q}}
\simeq
L_1\varepsilon_1+\ldots+L_m\varepsilon_m
-
(L_m+\mathfrak{q})\varepsilon_{n-m+1}
-\ldots-
(L_1+\mathfrak{q})\varepsilon_n\,,
\end{equation}
which reproduce our further result \eqref{MagneticYoungDiagramsGr}.
Here we have used the freedom to shift all coefficients of the $\varepsilon_i$ by the same constant, which does not change the corresponding $\mathfrak{su}(n)$ weight. In the present case, the required shift is $-\frac{\mathfrak{q}m}{n}$.

In this case, the quadratic Casimir eigenvalues for \eqref{MagnWeight} are modified according to
\begin{align}
\mathcal{C}_2(\rho_{\mathfrak{q}})
=
\langle\rho_{\mathfrak{q}},\rho_{\mathfrak{q}}+\delta\rangle&=
\mathcal{C}_2(\rho_0)
+
2\mathfrak{q}\langle\rho_0,\omega_{n-m}\rangle
+
\mathfrak{q}\langle\omega_{n-m},\delta\rangle
+
\mathfrak{q}^2\langle\omega_{n-m},\omega_{n-m}\rangle=
\nonumber\\
&=
\mathcal{C}_2(\rho_0)
+
\mathfrak{q}\sum\limits_{k=1}^m L_k
+
\frac{m(n-m)}{2}\mathfrak{q}
+
\frac{m(n-m)}{2n}\mathfrak{q}^2=
\nonumber\\
&=
\sum\limits_{k=1}^m
\bigg(L_k+\frac{\mathfrak{q}}{2}\bigg)
\bigg(L_k+\frac{\mathfrak{q}}{2}+n-2k+1\bigg)
+
\frac{m(n-2m)}{4n}\mathfrak{q}^2\,,\label{MagneticCasimirFormula}
\end{align}
where we used
\begin{equation}
2\langle\rho_0,\omega_{n-m}\rangle=\sum\limits_{k=1}^mL_k\,,\quad \langle\omega_{n-m},\delta\rangle=\frac{m(n-m)}{2}\,,\quad\langle\omega_{n-m},\omega_{n-m}\rangle=\frac{m(n-m)}{2n}\,.
\end{equation}

\section{First-order GLSM approach}\label{SectionGLSM}

Starting with this section, we turn to the new results of the paper. In particular, we present two alternative derivations of the spectrum~\eqref{MagneticCasimirFormula} and give an explicit construction of Grassmannian harmonics.

Here we discuss the method originally introduced in the context of two-dimensional Euclidean sigma models\footnote{This method may be viewed as a synthesis of the GLSM \cite{DAdda:1978vbw,Witten:1978bc} and the first-order \cite{Losev:2005pu,Gamayun:2023sif,Gamayun:2023atu} approaches to sigma models. Structurally, it takes the form of a current--$\overline{\text{current}}$ deformation coupling \cite{Gerganov:2000mt,Gamayun:2026zav} of holomorphic and antiholomorphic $\beta\gamma$ systems \cite{Nekrasov:2005wg}.} \cite{BykovNilp,BykovFerm,Bykov:2021,BykovRiem}. The idea behind this method, also known as the ``Gross\,--\,Neveu formulation'', is to find a more tractable equivalent description of a 2D sigma model with target space $\mathcal{M}$ by introducing momenta conjugate to the target-space coordinates, thereby passing to $\mathrm{T}^\ast\mathcal{M}$, and then realizing this space as a symplectic reduction (i.e.~a gauge system) of a flat space $\mathrm{T}^\ast\mathbb{C}^N$ with respect to an appropriate gauge group, whenever such a realization is possible, see \cite{Bykov2023}. For an introduction to these ideas, see \cite{Krivorol:2025tqj}. Within this framework, the corresponding first-order GLSM formulation for Grassmannian target spaces was constructed in \cite{BykovKrivorol2023}.

In this section, we study the 1D reduction of this method, i.e.~quantum mechanics on Grassmannian manifolds, whose quantization naturally leads to the spectral problem for the Laplacian. The advantage of this formulation is that, for Grassmannians $\mathsf{Gr}(m,n)$, it provides a natural redundant coordinate system which, as we will see, generalizes the usual homogeneous coordinates.

The basic geometric idea is not to work directly on the Grassmannian charts itself, which has several disadvantages. For instance, the natural $\mathbf{SU}(n)$-invariance of the normal metric is not manifest in local coordinates, but rather on the principal $\mathbf{GL}(m,\mathbb{C})$ Stiefel bundle $\mathsf{St}(m,n)\rightarrow\mathsf{Gr}(m,n)$. In other words, the redundant coordinates are provided by Stiefel frames, i.e.~sets of $m$ linearly independent vectors $u_1,\ldots,u_m$, assembled into a single matrix $U$ whose columns are the vectors $u_\alpha$. Different choices of a frame for the same $m$-dimensional subspace are related by the natural $\mathbf{GL}(m,\mathbb{C})$ action. We further enlarge this space by introducing the conjugate ``momentum directions'' $V$. In this first-order framework, the $\mathbf{GL}(m,\mathbb{C})$ gauge symmetry is implemented by the matrix-valued  constraint $VU=0$. The significance of this step is that one can now apply the methods of constrained-system quantization and obtain a remarkably simple Hamiltonian which, after fixing the $\mathbf{GL}(m,\mathbb{C})$ gauge, reproduces the corresponding Laplace operators.

This formulation not only greatly simplifies the computation of the Laplace\,--\,Beltrami and Laplace\,--\,Bochner eigenvalues, but also yields a simple and transparent representation of the corresponding eigenfunctions, or Grassmannian harmonics. This construction may be viewed as a natural generalization of the representation of spherical harmonics in terms of harmonic homogeneous polynomials restricted to the sphere; see \cite{Bykov2023}.

\subsection{Demonstration of the method: $\mathbb{CP}^{n-1}$ case.}\label{SectionCP}
\label{SectionCPnSpectrum}
For pedagogical purposes, we begin with the GLSM formulation of the spectral problem for the Laplace operator on $\mathbb{CP}^{n-1}$, presenting it in a form that will be generalized below to arbitrary Grassmannians. Much of this construction is already known, and a substantial part of the material in this section follows \cite[Section~2]{BykovSmilga}. We include it here in order to illustrate the main ideas and to fix the notation and conventions used throughout the rest of the paper.
\subsubsection{Without monopoles.}

Let us consider the so-called ``first-order GLSM'' action\footnote{Here and below, Latin indices run from $1$ to $n$. The position of indices and the use of bars are purely matters of notation and aesthetics; we assume that the precise geometric meaning of the corresponding indexed objects is clear from the context.}
\begin{equation}
\label{GNforCP}
\mathcal{S}[u,v,\mathcal{A}]=\int \mathrm{d}t\,
\bigg[\Big(v_i \dot{u}^i-i\thickbar{\mathcal{A}}\big(v_i u^i\big)-c.c.\Big) + |u|^2|v|^2 \bigg] ,
\end{equation}
which describes a first-order theory on the phase space $\mathrm{T}^\ast\mathbb{C}^n$ with first-class constraints. Here $u^i$ and $v_i$ are canonical coordinates on $\mathrm{T}^\ast\mathbb{C}^n$, the field $\thickbar{\mathcal{A}}$ is the Lagrange multiplier imposing $v_i u^i=0$, and $c.c.$ denotes the complex conjugate terms. It is a $\mathbb{C}^\ast$-gauge theory with respect to the local symmetry
\begin{equation}
u^i\rightarrow \lambda(t)u^i\,,
\qquad
v_i\rightarrow \lambda(t)^{-1}v_i\,,\qquad \lambda\in\mathbb{C}^\ast\,,
\end{equation}
and $\thickbar{\mathcal{A}}$ is the corresponding gauge field.

The main property of the action \eqref{GNforCP} is its equivalence \cite{Bykov:2021} to the one-dimensional~$\mathbb{CP}^{n-1}$ sigma model equipped with the Fubini\,--\,Study metric, whose  quantization gives the Laplace\,--\,Beltrami spectral problem with this metric. 
To establish this equivalence, one eliminates the auxiliary fields $v$ and $\mathcal{A}$ by their equations of motion and fixes the gauge, for instance by passing to inhomogeneous coordinates $u^1=1$. Thus, the canonical quantization of \eqref{GNforCP} is equivalent to the quantization of the sigma model. This result explains the name ``first-order gauged linear sigma model (GLSM)''. Indeed, in order to linearize the second-order theory  we introduce the momenta $v$ conjugate to $u$, the gauge field $\thickbar{\mathcal{A}}$, and write the theory in the first-order form \eqref{GNforCP}. 

The advantage of the first-order GLSM formulation is that the quantization of \eqref{GNforCP} gives a simpler description of both the Hamiltonian and the Hilbert space than the direct quantization of the sigma model. We choose the canonical commutation relations in the form\footnote{The absence of the factor $i$ in the canonical commutation relations is due to our convention that the kinetic term in the action is purely imaginary. This can be viewed as a one-dimensional analogue of the Euclidean formalism, and is also reflected in the sign of the Hamiltonian in \eqref{GNforCP}.}
\begin{equation}
[u^i,v_j]=\delta^i_j\,, \qquad [\bar u^k,\bar v_k]=-\delta^{\bar{k}}_{\bar{q}}\,.
\end{equation}
They are realized by
\begin{equation}\label{CanonicalQuantizationCP}
v_i\mapsto -\partial_i:=-\frac{\partial}{\partial u^i},
\qquad
\bar v_i\mapsto \smallthickbar{\partial}_{i}:=\frac{\partial}{\partial\bar u^i}.
\end{equation}
Thus the Hamiltonian becomes
\begin{equation}
\label{HamiltonianCP}
\mathsf H=|u|^2|v|^2
\quad\Longrightarrow\quad
\hat{\mathsf H}\psi
=
-|u|^2\delta_{k\bar{q}}\frac{\partial^2\psi}{\partial u_k\partial\bar u_q}
=
-|u|^2\partial_k\smallthickbar{\partial}_k\psi\,.
\end{equation}

It remains to determine the physical Hilbert space. Following the Dirac prescription for systems with first-order constraints, we impose the constraints on states\footnote{The ordering in the quantization of the constraints is chosen in such a way that they determine functions on projective space.  In turn, these constraints substantially reduce the ordering ambiguity in the Hamiltonian~(\ref{HamiltonianCP}). Indeed, ambiguities in the ordering of $u$'s and $\partial$'s (and their conjugates) are either constants or are proportional to the constraints and hence vanish.}:
\begin{equation}
\label{ConstraintAsPdeCP}
u^i v_i=0\,,
\qquad
\bar u^i\bar v_i=0
\quad\Longrightarrow\quad
u^i\frac{\partial\psi_{\mathrm{phys}}}{\partial u^i}=\bar u^i\frac{\partial\psi_{\mathrm{phys}}}{\partial\bar u^i}=0\,.
\end{equation}
Hence physical states are homogeneous of degree zero separately in $u^i$ and $\bar u^i$. The next step is to choose a convenient basis in this space. Since the theory \eqref{GNforCP} has a global $\mathbf{SU}(n)$-symmetry, $u\mapsto gu$, $v\mapsto vg^{-1}$, with $g\in\mathbf{SU}(n)$, it is natural to organize the basis in terms of~$\mathbf{SU}(n)$ representations. The equivalence between the first-order GLSM \eqref{GNforCP} and the sigma model implies that the spectral problem for $\hat{\mathsf{H}}$ on the physical Hilbert space is equivalent to the spectral problem for $-\triangle_{\mathbb{CP}^{n-1}}$ on $L^2(\mathbb{CP}^{n-1})$.

The simplest gauge-invariant homogeneous function built from one $u$ and one $\bar u$ is
\begin{equation}
\label{psi1CP}
\psi_1 = \frac{T_{k\bar q}\,u^k\bar u^q}{|u|^2}\,,
\qquad
T_{k\bar q}\in \mathscr V\otimes\thickbar{\mathscr{V\,}}\,.
\end{equation}
Here $\mathscr V$ is the defining representation of $\mathbf{SU}(n)$ and $\thickbar{\mathscr V\,}$ is its conjugate. Let us determine the additional condition on $T$ under which it becomes an eigenfunction. Using \eqref{HamiltonianCP}, a direct computation gives
\begin{equation}
\hat{\mathsf H}\psi_1
=
n\psi_1-\delta^{k\bar q}T_{k\bar q}\,.
\end{equation}
Thus it is enough to require $T$ to be traceless.
The same mechanism works in general. Namely, one has the following statement, which is Theorem 1 of \cite{BykovSmilga}.\\
\textbf{Proposition.} The functions of the form\footnote{Historically, such wavefunctions were obtained in \cite{Kuwabara1988SPECTRUMOT} by a different method.}
\begin{equation}
\label{EigenfunctionsCP}
\psi_L
=
\frac{
T_{k_1\ldots k_L\bar q_1\ldots \bar q_L}
u^{k_1}\ldots u^{k_L}
\bar u^{q_1}\ldots \bar u^{q_L}
}{
|u|^{2L}
}\,,
\qquad
T_{k_1\ldots k_L\bar q_1\ldots \bar q_L}
\in
\Big(\mathscr V\otimes\thickbar{\mathscr V}\Big)^{\otimes L}
\end{equation}
are eigenfunctions of the Hamiltonian \eqref{HamiltonianCP} with the eigenvalue
\begin{empheq}[box = \fcolorbox{black}{white}]{align}
E_L^{\mathbb{CP}^{n-1}} = L(L+n-1)\,.\label{CPspectrum}
\end{empheq}
if the tensor $T$ is traceless with respect to any holomorphic-antiholomorphic pair of indices.
The proof is a direct computation. This agrees with the standard spectrum of~$\mathbb{CP}^{n-1}$. 

It remains to identify the irreducible $\mathbf{SU}(n)$-representations carried by these eigenfunctions. In terms of Young diagrams, each copy of $\mathscr{V}$ in \eqref{EigenfunctionsCP} corresponds to one box, while each copy of the conjugate representation $\thickbar{\mathscr{V}\,}$ corresponds to a column of height $n-1$. Since the variables $u^i$ commute, not all components of $\mathscr{V}^{\otimes L}$ contribute to the numerator in \eqref{EigenfunctionsCP}. More precisely, consider the contraction map
\begin{equation}
\mathcal{C}_L: \mathscr{V}^{\otimes L} \longrightarrow \mathbb{C}[u]_L\,,
\qquad
T_{i_1\ldots i_L} \longmapsto T_{i_1\ldots i_L}u^{i_1}\cdots u^{i_L}\,,
\end{equation}
where $\mathbb{C}[u]_L$ denotes homogeneous polynomials of degree $L$ in $u$. Its image is the $L$\,-th symmetric power $\operatorname{Sym}^L\mathscr{V}$, while its kernel consists of tensors which vanish after contraction with the commuting variables $u^i$. In terms of Young diagrams this gives\footnote{Here and below, colors in Young diagrams are used only for bookkeeping.}
\begin{equation}
\underbrace{
\begin{ytableau}
*(lime)
\end{ytableau}
\,\otimes\,
\begin{ytableau}
*(lime)
\end{ytableau}
\,\otimes\,
\dots
\,\otimes\,
\begin{ytableau}
*(lime)
\end{ytableau}
}_{L\text{ times}} \quad\simeq\quad
\underbrace{
\begin{ytableau}
*(lime) & *(lime) & *(lime) & *(lime) & *(lime) & *(lime) & *(lime)
\end{ytableau}
}_{L\text{ boxes}}
~\oplus~
\operatorname{Ker}\mathcal{C}_L\,.
\end{equation}
Equivalently,
\begin{equation}
\operatorname{Im}\mathcal{C}_L
\simeq
\operatorname{Sym}^L\mathscr{V}
\simeq
\underbrace{
\begin{ytableau}
*(lime) & *(lime) & *(lime) & *(lime) & *(lime) & *(lime) & *(lime)
\end{ytableau}
}_{L\text{ boxes}}\,.
\end{equation}
The antiholomorphic variables give the conjugate symmetric representation $\operatorname{Sym}^L\thickbar{\mathscr{V}\,}$.
Hence the relevant tensor product is
\begin{equation}
\operatorname{Sym}^L\mathscr{V} \otimes \operatorname{Sym}^L\thickbar{\mathscr{V}\,}\,.
\end{equation}
The trace components in this tensor product correspond to lower-level eigenfunctions. Removing them selects the irreducible representation with Young diagram $\mathbb{Y}(L)$ of the form
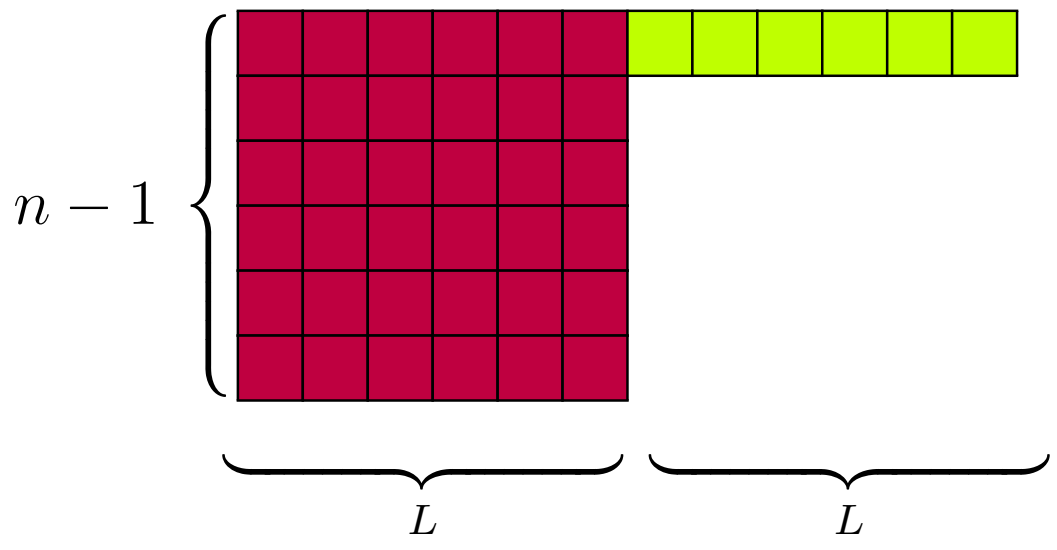
\begin{equation}
\label{YoungDiagrCP}
\vcenter{
\hbox{
$
\begin{array}{@{}c@{}}
\vcenter{\hbox{$n-1$}}
\,
\left\{
\begin{ytableau}
*(purple) & *(purple) & *(purple) & *(purple) & *(purple) & *(purple) & *(lime) & *(lime) & *(lime) & *(lime) & *(lime) & *(lime)\\
*(purple) & *(purple) & *(purple) & *(purple) & *(purple) & *(purple)\\
*(purple) & *(purple) & *(purple) & *(purple) & *(purple) & *(purple)\\
*(purple) & *(purple) & *(purple) & *(purple) & *(purple) & *(purple)\\
*(purple) & *(purple) & *(purple) & *(purple) & *(purple) & *(purple)\\
*(purple) & *(purple) & *(purple) & *(purple) & *(purple) & *(purple)
\end{ytableau}
\right.
\\[-0.3em]
\hspace*{1.55cm}
\underbrace{\hspace{2.70cm}}_{L}
\hspace*{0.05cm}
\underbrace{\hspace{2.70cm}}_{L}
\end{array}
$
}
}
\end{equation}
In this diagram, the green boxes come from $\mathrm{Sym}^L\mathscr{V}$, while the red rectangle comes from~$\mathrm{Sym}^L\thickbar{\mathscr{V}\,}$.
Since the eigenfunctions of the Laplace operator form a basis in $L^2\big(\mathbb{CP}^{n-1}\big)$, we obtain the decomposition
\begin{equation}
L^2\big(\mathbb{CP}^{n-1}\big)
\simeq
\bigoplus_{L=0}^{\infty}
\mathbb{Y}(L)\,,
\end{equation}
where $\mathbb{Y}(L)$ denotes the irreducible $\mathbf{SU}(n)$-representation labeled by the corresponding Young diagram. 

\subsubsection{Monopole harmonics.}\label{SectionCPharmonics}

Let us generalize the construction above to the case of a magnetic monopole. In other words, we modify the first-order GLSM in order to describe the spectral problem associated with the Laplace\,--\,Bochner operator and to construct the corresponding monopole harmonics. This problem was considered in \cite[Section~2.3.2]{BykovSmilga}. As explained there, the magnetic monopole field is incorporated by modifying the constraint in \eqref{GNforCP} as
\begin{equation}\label{MagneticConstraintsCP}
v_i u^i = 0
\qquad\mapsto\qquad
v_i u^i = \frac{\mathfrak{q}}{2}
\end{equation}
together with its complex conjugate, where $\mathfrak{q}$ has the meaning of the magnetic monopole charge\footnote{For simplicity, we assume $\mathfrak{q}\geq0$.}. Eliminating the auxiliary fields $v$ and $\mathcal{A}$ and fixing the gauge then gives the $\mathbb{CP}^{n-1}$ sigma model coupled to the corresponding magnetic field of monopole charge $\mathfrak{q}$, as follows by direct computation.

Quantizing the constraint \eqref{MagneticConstraintsCP} and its complex conjugate according to \eqref{CanonicalQuantizationCP} gives the physical-state conditions
\begin{equation}
u^i\frac{\partial\psi_{\mathrm{phys}}}{\partial u^i}
=
-\frac{\mathfrak{q}}{2}\psi_{\mathrm{phys}}\,,
\qquad
\bar u^i\frac{\partial\psi_{\mathrm{phys}}}{\partial\bar u^i}
=
\frac{\mathfrak{q}}{2}\psi_{\mathrm{phys}}\,,
\end{equation}
which generalize \eqref{ConstraintAsPdeCP}. 

The natural generalization of the ansatz \eqref{EigenfunctionsCP} is\footnote{These wavefunctions satisfy
\begin{equation}
\psi^{\mathfrak{q}}_L\big(e^{i\varphi}u\big)
=
e^{-i\mathfrak{q}\varphi}
\psi^{\mathfrak{q}}_L\big(u\big)\,,
\end{equation}
and hence are $\mathbf{U}(1)$-equivariant functions of weight $-\mathfrak{q}$ on the Hopf bundle $\mathbb{S}^{2n-1}\rightarrow\mathbb{CP}^{n-1}$. They may therefore be identified with sections of the standard line bundle $\mathscr{O}(-\mathfrak{q})$.}
\begin{equation}\label{EigenfunctionsCPq}
\psi_{L}^{\mathfrak{q}}
=
\frac{
T_{k_1\ldots k_L\bar q_1\ldots \bar q_{L+\mathfrak{q}}}
u^{k_1}\ldots u^{k_L}
\bar u^{q_1}\ldots \bar u^{q_{L+\mathfrak{q}}}
}{
|u|^{2L+\mathfrak{q}}
}\,,
\end{equation}
where $T$ still satisfies the tracelessness condition. Here we impose the standard quantization condition $\mathfrak{q}\in\mathbb{Z}$. Acting with the first-order GLSM Hamiltonian \eqref{HamiltonianCP} on the wavefunctions \eqref{EigenfunctionsCPq}, we obtain
\begin{empheq}[box = \fcolorbox{black}{white}]{align}\label{CPspectrumMagnetic}
E_{L;\mathfrak{q}}^{\mathbb{CP}^{n-1}}
=
\bigg(L+\frac{\mathfrak{q}}{2}\bigg)
\bigg(L+\frac{\mathfrak{q}}{2}+n-1\bigg)\,.
\end{empheq}
Thus, at the level of the first-order GLSM Hamiltonian, the effect of the magnetic monopole is simply the shift
$L\mapsto L+\frac{\mathfrak{q}}{2}$ in the $\mathbb{CP}^{n-1}$ spectrum \eqref{CPspectrum}. Note, however, that this expression differs from \cite[formula~(2.32)]{BykovSmilga} by the constant shift $\frac{\mathfrak{q}^2}{4}$. This constant accounts for the difference between the first-order GLSM Hamiltonian and the Laplace\,--\,Bochner operator.

Also, using the same technique as in the previous section, one can show that the Young diagrams corresponding to the wavefunctions of the form \eqref{EigenfunctionsCPq} are
\begin{equation}
\vcenter{
\hbox{
$
\begin{array}{@{}c@{}}
\vcenter{\hbox{$n-1$}}
\,
\left\{
\begin{ytableau}
*(purple) & *(purple) & *(purple) & *(purple) & *(purple) & *(purple) &
*(orange) & *(orange) & *(orange) &
*(lime) & *(lime) & *(lime) & *(lime) & *(lime) & *(lime)\\
*(purple) & *(purple) & *(purple) & *(purple) & *(purple) & *(purple) &
*(orange) & *(orange) & *(orange)\\
*(purple) & *(purple) & *(purple) & *(purple) & *(purple) & *(purple) &
*(orange) & *(orange) & *(orange)\\
*(purple) & *(purple) & *(purple) & *(purple) & *(purple) & *(purple) &
*(orange) & *(orange) & *(orange)\\
*(purple) & *(purple) & *(purple) & *(purple) & *(purple) & *(purple) &
*(orange) & *(orange) & *(orange)\\
*(purple) & *(purple) & *(purple) & *(purple) & *(purple) & *(purple) &
*(orange) & *(orange) & *(orange)
\end{ytableau}
\right.
\\[-0.3em]
\hspace*{1.60cm}
\underbrace{\hspace{2.6cm}}_{L}
\hspace*{0.02cm}
\underbrace{\hspace{1.25cm}}_{\mathfrak{q}}
\hspace*{0.02cm}
\underbrace{\hspace{2.6cm}}_{L}
\end{array}
$
}
}
\end{equation}

\subsection{Unitary Grassmannians.}\label{SectionGr}
Motivated by the example of $\mathbb{CP}^{n-1}$, we now turn to the spectral problem for the unitary Grassmannian\footnote{Since $\mathsf{Gr}(m,n)\simeq\mathsf{Gr}(n-m,n)$, throughout this section we assume $m\leq n-m$.} $\mathsf{Gr}(m,n)$. Our aim is to reproduce the results \eqref{SphericalCasimirGr} and \eqref{MagneticCasimirFormula} within the first-order formalism. The same spectral problem was previously solved using branching rules for $\mathbf{SU}(n)$ in \cite{BenHalima}. Here we derive these results by a direct generalization of the first-order construction developed in the previous subsection.

\subsubsection{Without magnetic field.}
The natural matrix analogue of \eqref{GNforCP} for $\mathsf{Gr}(m,n)$ is the action \cite{BykovKrivorol2023}
\begin{equation}
\label{GNforGr}
\mathcal{S}[U,V,\mathcal{A}]
=
\int\mathrm{d}t\,
\Big[
\mathrm{Tr}\,\big(V\dot{U}-i\thickbar{\mathcal{A}}\,(VU)-c.c.\big)
+
\mathrm{Tr}\big(\thickbar{U}UV\thickbar{V}\big)
\Big]\,.
\end{equation}
Here
\begin{equation}
U\in \operatorname{Mat}_{n\times m}(\mathbb C)\,,\qquad
V\in \operatorname{Mat}_{m\times n}(\mathbb C)\,,\qquad
\thickbar{\mathcal{A}}\in \mathfrak{gl}(m,\mathbb C)\,,
\end{equation}
where we assume $\mathrm{rank}(U) = m$. 
The action is invariant under the local gauge \(\mathbf{GL}(m,\mathbb C)\)-transformations
\begin{equation}\label{GLgaugeGroup}
U\mapsto Ug,\qquad
V\mapsto g^{-1}V,\qquad
\thickbar{\mathcal{A}}\mapsto g^{-1}\thickbar{\mathcal{A}}g-i g^{-1}\dot g \,,
\qquad g(t)\in \mathbf{GL}(m,\mathbb C)\,.
\end{equation}
Variation with respect to $\thickbar{\mathcal{A}}$ imposes the constraint $VU=0$. After eliminating the auxiliary fields $V,\mathcal{A}$ and fixing the gauge freedom, this action reduces to the sigma model action on~$\mathsf{Gr}(m,n)$ equipped with the natural invariant metric, the Grassmannian analogue of the Fubini\,--\,Study metric, see \cite{BykovKrivorol2023} for details. 

Thus, to analyze the spectral problem for the Hamiltonian, we canonically quantize~\eqref{GNforGr}. In terms of matrix components, the canonical commutation relations are\footnote{Here and below, Greek indices run from $1$ to $m$.}
\begin{equation}
\big[U_{i\alpha},V_{\beta j}\big]
=
\delta_{ij}\delta_{\alpha\beta}\,,
\qquad
\big[\,\thickbar{U}_{\alpha i},\thickbar{V}_{j\beta}\big]
=
-\delta_{ij}\delta_{\alpha\beta}\,.
\end{equation}
They are realized in the coordinate representation as
\begin{equation}
\label{CCR}
V_{\alpha i}\longrightarrow -\partial_{\alpha i},
\qquad
\thickbar{V}_{i\alpha}\longrightarrow \thickbar{\partial}_{i\alpha},
\qquad\text{where}\qquad
\partial_{\alpha i}:=\frac{\partial}{\partial U_{i\alpha}},
\qquad
\thickbar{\partial}_{i\alpha}:=\frac{\partial}{\partial \thickbar{U}_{\alpha i}}.
\end{equation}
The classical Hamiltonian associated with \eqref{GNforGr} is
$\mathsf{H}=\mathrm{Tr}\big(\thickbar{U}UV\thickbar{V}\big)$.
In the representation \eqref{CCR}, the corresponding quantum Hamiltonian takes the form
\begin{equation}\label{HamiltonianGr}
\hat{\mathsf{H}}
=
-\big(\thickbar{U}U\big)_{\alpha\beta}
{\partial}_{\beta i}\thickbar{\partial}_{i\alpha}\,,
\end{equation}
and, upon restriction to the physical states, reproduces minus the Laplace operator on the Grassmannian.

The next step is to construct gauge-invariant functions in the physical sector of the Hilbert space. In direct analogy with \eqref{psi1CP}, the simplest non-trivial example is
\begin{equation}
\psi_{1}
=
T_{k\bar{q}}
\Big(U\big(\thickbar{U}U\big)^{-1}\thickbar{U}\Big)^{k\bar q}\,,
\end{equation}
where, as in the previous section, $T_{k\bar{q}}$ is a tensor in $\mathscr{V}\otimes\thickbar{\mathscr{V}\,}$, with $\mathscr{V}$ the defining representation of $\mathbf{SU}(n)$ and $\thickbar{\mathscr{V}\,}$ its complex conjugate. This naturally suggests the following ansatz for higher eigenfunctions:
\begin{equation}
\label{AnsatzGr}
\psi_L
=
T_{a\ldots k \;\bar{b}\ldots \bar{q}}
\Big(U\big(\thickbar{U}U\big)^{-1}\thickbar{U}\Big)^{a\bar b}
\ldots
\Big(U\big(\thickbar{U}U\big)^{-1}\thickbar{U}\Big)^{k\bar q}\,,
\qquad
T_{a\ldots k \;\bar{b}\ldots \bar{q}}
\in
\Big(\mathscr{V}\otimes\thickbar{\mathscr{V}\,}\Big)^{\otimes L}\,.
\end{equation}
Within this formalism, the Grassmannian eigenfunctions can be represented in the form~\eqref{AnsatzGr}, subject to suitable conditions on the tensor $T$. It therefore remains to determine these conditions. As in the case of $\mathbb{CP}^{n-1}$, one first requires $T$ to be traceless with respect to any pair of indices of different type. For complex Grassmannians, however, one must additionally diagonalize a finite-dimensional operator acting on the antiholomorphic indices of $T$.

Also note that the appearance of the gauge-invariant combination of $U$ and $\thickbar{U}$ in \eqref{AnsatzGr} has a direct geometric meaning. Indeed,
\begin{equation}
P:=U\big(\thickbar{U}U\big)^{-1}\thickbar{U}
\end{equation}
is the Hermitian projector onto the $m$-dimensional subspace spanned by the columns of~$U$. Hence the Grassmannian itself may be identified with the space of rank-$m$ Hermitian projectors \cite{harris},
\begin{equation}
\mathsf{Gr}(m,n)
\simeq
\big\{
P\in\mathrm{End}(\mathbb{C}^n)
\ \big|\
P^2=P,\quad P^\dagger=P,\quad \mathrm{Tr}(P)=m
\big\}\,.
\end{equation}
Thus, the matrix elements $P^{k\bar{q}}$ provide natural global coordinates on the Grassmannian, and the ansatz \eqref{AnsatzGr} realizes Grassmannian harmonics as polynomials in these projector coordinates. 
We also note that, in terms of the familiar \textit{Pl\"ucker coordinates} (see \cite{harris})
\begin{align}\label{Plucker}
   \Pi^{b_1\ldots b_m}
:=
\varepsilon^{\alpha_1\ldots\alpha_m}
U_{\alpha_1}^{b_1}\ldots
U_{\alpha_m}^{b_m} \,,
\end{align}
the projector $P$ may be expressed as
\begin{align}
    P^{a\bar b}\sim \frac{\Pi^{a k_1\ldots k_{m-1}}\thickbar{\Pi}^{\,\bar b\,\bar k_1\ldots\bar k_{m-1}}}{\mathrm{det}\big(\thickbar{U} U\big)}\,.
\end{align}

We now determine the possible energy eigenvalues carried by the ansatz \eqref{AnsatzGr}. To this end, let us introduce the following ``spin-chain'' operator:
\begin{equation}
\label{HamiltonianForTensors}
\mathscr{J}
=
nL\,\mathds{1}
+
\sum_{\substack{i,j=1\\ i\neq j}}^{L}
\thickbar{\mathbb{P}}_{ij}
~\in~
\mathrm{End}\Big(\thickbar{\mathscr{V}\,}^{\otimes L}\Big)\,.
\end{equation}
Here $\mathds{1}$ is the identity operator, while $\thickbar{\mathbb{P}}_{ij}$ is the transposition operator acting on the $i$-th and $j$-th copies of $\thickbar{\mathscr{V}\,}$ in the tensor product. Accordingly, $\thickbar{\mathbb{P}}_{ij}$ acts on the tensor $T_{a\ldots k \;\bar{b}\ldots \bar{q}}$ by permuting the corresponding antiholomorphic indices. For instance,
\begin{equation}
\thickbar{\mathbb{P}}_{12}T_{a b \bar c\bar d}
=
T_{a b \bar d\bar c}\,.
\end{equation}

\textbf{Proposition.} Suppose that $T$ is traceless with respect to any pair of indices of different type. Then the action of the Hamiltonian on the function \eqref{AnsatzGr} reduces to the action of the operator $\mathscr{J}$ on the antiholomorphic indices of $T$:
\begin{align}
\hat{\mathsf{H}}\psi_L
=&
\big(\mathscr{J}T_{a\ldots k \;\bar{b}\ldots \bar{q}}\big)
\Big(U\big(\thickbar{U}U\big)^{-1}\thickbar{U}\Big)^{a\bar b}
\ldots
\Big(U\big(\thickbar{U}U\big)^{-1}\thickbar{U}\Big)^{k\bar q}
=\nonumber\\
=&
nL\psi_L
+
\sum_{\substack{i,j=1\\ i\neq j}}^{L}
\Big(\thickbar{\mathbb{P}}_{ij}T_{a\ldots k \;\bar{b}\ldots \bar{q}}\Big)
\Big(U\big(\thickbar{U}U\big)^{-1}\thickbar{U}\Big)^{a\bar b}
\ldots
\Big(U\big(\thickbar{U}U\big)^{-1}\thickbar{U}\Big)^{k\bar q}\,.
\label{SpinCHainAction}
\end{align}
The proof of this statement follows by direct differentiation. Consequently, within the ansatz~\eqref{AnsatzGr}, the spectral problem for $\hat{\mathsf{H}}$ reduces to the finite-dimensional spectral problem for the ``spin-chain'' Hamiltonian $\mathscr{J}$ acting on the antiholomorphic tensor factors $\thickbar{\mathscr{V}\,}^{\otimes L}$.

We assume that the antiholomorphic indices are symmetrized according to a Young diagram of the general form\footnote{We also assume that the holomorphic indices are symmetrized according to the Young diagram of the representation conjugate to \eqref{AntiHolomorphicYoungDiagramGr}.}
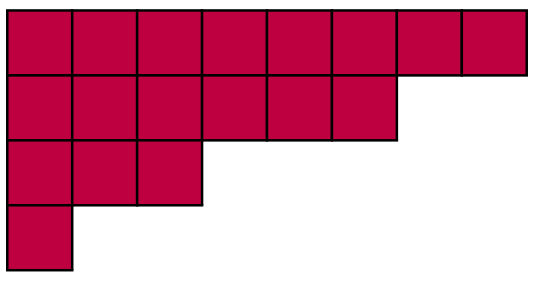
\begin{equation}
\label{AntiHolomorphicYoungDiagramGr}
\begin{ytableau}
*(purple) & *(purple) & *(purple) & *(purple) & *(purple) & *(purple) & *(purple) & *(purple) \\
*(purple) & *(purple) & *(purple) & *(purple) & *(purple) & *(purple) \\
*(purple) & *(purple) & *(purple) \\
*(purple)
\end{ytableau}
\end{equation}
Let us first observe that tensors symmetrized according to Young diagrams of height greater than $m$ lie in the kernel of the contraction map\footnote{By the kernel of the contraction map we mean the subspace of tensors that give $\psi_L=0$ upon substitution into \eqref{AnsatzGr}.}. 
Indeed, the projector $P$ has rank $m$, so every $r\times r$ minor with $r>m$ vanishes: 
\begin{equation} \sum_{\sigma\in \mathbf{S}_r}(-1)^\sigma P^{a_1\bar b_{\sigma(1)}}\ldots P^{a_r\bar b_{\sigma(r)}}=\det\Big[P^{a_i\bar b_j}\Big]_{i,j=1}^{r}=0\,. 
\end{equation}
The sum on the left antisymmetrizes the $r$ antiholomorphic indices, precisely as required by a column of $r$ boxes in the Young diagram for the antiholomorphic indices of $T$.
Therefore, only Young diagrams of height at most $m$ contribute. We denote such diagrams by $\mathbb{Y}(L_1,\ldots,L_m)$, where $L_i$ is the length of the $i$-th row. Taking this into account, we can formulate the main statement of this section.

\textbf{Proposition.} The eigenvalues of the Hamiltonian $\hat{\mathsf H}$ are
\begin{empheq}[box = \fcolorbox{black}{white}]{align}
\label{SpectrumGr}
E^{\mathsf{Gr}(m,n)}_{L_1,\ldots,L_m}
=
\sum_{j=1}^m
L_j\big(L_j+n-2j+1\big)\,,
\end{empheq}
where $L_1\geq L_2\geq\cdots\geq L_m\geq0$ are integers specifying the Young diagram $\mathbb{Y}(L_1,\ldots,L_m)$.

\begin{proof}
By \eqref{SpinCHainAction}, it is enough to diagonalize the sum of all transpositions
\begin{equation}
\thickbar{\mathbb{P}}
:=
\sum_{\substack{i,j=1\\ i\neq j}}^{L}
\thickbar{\mathbb{P}}_{ij}
~\in~
\mathrm{End}\Big(\thickbar{\mathscr{V}\,}^{\otimes L}\Big)\,.
\end{equation}
It is useful to regard $\thickbar{\mathbb{P}}$ as an element of the group algebra $\mathbb{C}[\mathbf{S}_L]$. In this interpretation, it can be written as
\begin{equation}
\thickbar{\mathbb{P}}
=
2\sum\limits_{k=1}^L X_k\,,
\qquad
X_k
=
\sum\limits_{i=1}^{k-1}
\thickbar{\mathbb{P}}_{ik}
~\in~
\mathbb{C}[\mathbf{S}_L]\,,
\qquad
[X_i,X_j]=0\,,
\end{equation}
where the elements $X_k$ are the \textit{Jucys\,--\,Murphy elements} \cite{Jucys1974SymmetricPolynomials,Murphy1981YoungSeminormal,Okounkov1996}. They generate the maximal commutative subalgebra of $\mathbb{C}[\mathbf{S}_L]$ known as the \textit{Gelfand\,--\,Tsetlin subalgebra}.

The advantage of this representation of $\thickbar{\mathbb{P}}$ is that the operators $X_k$ are simultaneously diagonalizable on irreducible representations of $\mathbf{S}_L$, i.e.~on Specht modules \cite{WiltshireGordonWooZajaczkowska2017}. Their common eigenvectors form the \textit{Gelfand\,--\,Tsetlin basis}. In the Specht module associated with a Young diagram $\mathbb{Y}$ containing $L$ boxes, the basis vectors $\mathsf{v}_{\mathbb{T}}$ are labeled by standard Young tableaux $\mathbb{T}$ of shape $\mathbb{Y}$. The corresponding eigenvalues are \cite{Jucys1971Factorization}
\begin{equation}
X_k\mathsf{v}_{\mathbb{T}}
=
\big(j[k]-i[k]\big)\mathsf{v}_{\mathbb{T}}\,,
\end{equation}
where $i[k]$ and $j[k]$ are, respectively, the row and column coordinates of the box containing $k$ in the tableau $\mathbb{T}$. It follows that
\begin{equation}
\thickbar{\mathbb{P}}\,\mathsf{v}_{\mathbb{T}}
=
2\sum\limits_{k=1}^L
\big(j[k]-i[k]\big)\mathsf{v}_{\mathbb{T}}
=
2\sum\limits_{\Box\in\mathbb{Y}}
\big(j[\Box]-i[\Box]\big)\mathsf{v}_{\mathbb{T}}\,,
\end{equation}
since summing over the entries $k$ of a standard Young tableau is equivalent to summing over all boxes of its Young diagram.
For $\mathbb{Y}=\mathbb{Y}(L_1,\ldots,L_m)$, with $L=\sum_{i=1}^mL_i$, this gives
\begin{align}
2\sum\limits_{\Box\in\mathbb{Y}}
\big(j[\Box]-i[\Box]\big)
&=
2\sum\limits_{i=1}^m\sum\limits_{j=1}^{L_i}(j-i)
=
2\sum\limits_{i=1}^m
\Bigg(\frac{L_i(L_i+1)}{2}-iL_i\Bigg)=
\nonumber\\
&=
\sum\limits_{i=1}^m
L_i\big(L_i-2i+1\big)\,.
\end{align}
Finally, adding the contribution $nL\psi_L$ in \eqref{SpinCHainAction} and using $L=\sum_{i=1}^mL_i$, we obtain precisely~\eqref{SpectrumGr}.
\end{proof}

Thus, we reproduce the spectrum \eqref{SphericalCasimirGr}. The last ingredient is to determine the analogue of \eqref{YoungDiagrCP}, i.e.~the full Young diagram defining the eigenfunctions of the form \eqref{AnsatzGr}. So far, we only know that the anti-holomorphic indices can carry the symmetry \eqref{AntiHolomorphicYoungDiagramGr} and that the entire wavefunction must be traceless with respect to every holomorphic\,--\,anti-holomorphic pair of indices. Let us identify the corresponding irreducible representation in $\mathbb{Y}\otimes\thickbar{\mathbb{Y}}$, where~$\thickbar{\mathbb{Y}}$ denotes the representation complex conjugate to $\mathbb{Y}$.

The key observation is the following. The $\mathbf{SU}(n)$ representation associated with the Young diagram $\mathbb{Y}(L_1,\ldots,L_m)$ has highest weight represented by $\big(L_1,\ldots,L_m,0,\ldots,0\big)$ whereas the complex-conjugate representation
$\thickbar{\mathbb{Y}}(L_1,\ldots,L_m)$ has highest weight represented by $\big(0,\ldots,0,-L_m,\ldots,-L_1\big)$.
Here and below, an $n$-tuple $(a_1,\ldots,a_n)$ represents the weight
$\sum_{i=1}^n a_i\varepsilon_i$ upon restriction to the traceless Cartan subalgebra of $\mathfrak{su}(n)$. Accordingly, two such tuples differing by a common shift
$(a_1,\ldots,a_n)\mapsto(a_1+c,\ldots,a_n+c)$ represent the same $\mathfrak{su}(n)$ weight.
It follows from \cite[Corollary~2.3.1]{KOIKE198957} that the tensor product of representations 
$\mathbb{Y}(L_1,\ldots,L_m)\otimes\thickbar{\mathbb{Y}}(L_1,\ldots,L_m)$
contains a unique irreducible component obtained without contractions. Its highest weight is represented by
\begin{equation}\label{HighestWeightSU}
\big(
L_1,\ldots,L_m,
\underbrace{0,\ldots,0}_{n-2m},
-L_m,\ldots,-L_1
\big)\,.
\end{equation}
Adding $L_1$ to all components, which does not change the corresponding $\mathfrak{su}(n)$ weight, we obtain the equivalent representative
\begin{equation}\label{GrassmannianYoungWeights}
\big( 2L_1,L_1+L_2,\ldots,L_1+L_m, \underbrace{L_1,\ldots,L_1}_{n-2m}, L_1-L_m,L_1-L_{m-1},\ldots,0 \big)\,.
\end{equation}
If the coefficients $a_i$ in a representative $\sum_{i=1}^n a_i\varepsilon_i$ are non-negative integers satisfying the condition $a_1\geq a_2\geq\cdots\geq a_n=0$, they can be identified with the row lengths of the corresponding Young diagram. In the present case, this gives the $\mathbf{SU}(n)$ Young diagram with the lengths \eqref{GrassmannianYoungWeights}.
It admits a particularly transparent pictorial representation. The conjugate representation $\thickbar{\mathbb{Y}}$ is the complement of $\mathbb{Y}$ in the rectangle of height $n$, for example\footnote{Here we picture the particular example $\mathsf{Gr}(4,9)$.}
\begin{equation}
\begin{ytableau}
*(purple) & *(purple) & *(purple) & *(purple) &*(purple) &*(purple) &*(purple) &*(purple)  \\
*(purple) & *(purple) & *(purple) & *(purple) & *(purple) & *(purple) &*(lime) &*(lime) \\
*(purple) & *(purple) & *(purple) &*(lime) &*(lime) &*(lime) &*(lime) &*(lime) \\
*(purple) &*(lime) &*(lime) &*(lime) &*(lime) &*(lime) &*(lime) &*(lime) \\
*(lime) &*(lime) &*(lime) &*(lime) &*(lime) &*(lime) &*(lime) &*(lime) \\
*(lime) &*(lime) &*(lime) &*(lime) &*(lime) &*(lime) &*(lime) &*(lime) \\
*(lime) &*(lime) &*(lime) &*(lime) &*(lime) &*(lime) &*(lime) &*(lime) \\
*(lime) &*(lime) &*(lime) &*(lime) &*(lime) &*(lime) &*(lime) &*(lime) \\
*(lime) &*(lime) &*(lime) &*(lime) &*(lime) &*(lime) &*(lime) &*(lime)
\end{ytableau}
\end{equation}
And thus, the diagram with the lengths \eqref{GrassmannianYoungWeights} has the form
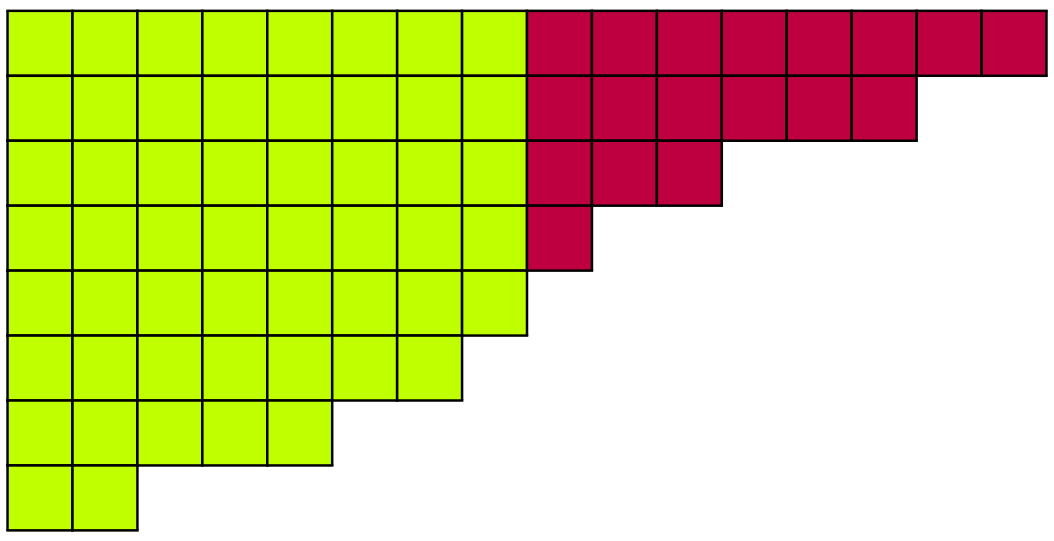
\begin{equation}
\label{DecompositionL2Gr}
\begin{ytableau}
*(lime) &*(lime) &*(lime) &*(lime) &*(lime) &*(lime) &*(lime) &*(lime) & *(purple) & *(purple) & *(purple) & *(purple) &*(purple) &*(purple) &*(purple) &*(purple) \\
*(lime) &*(lime) &*(lime) &*(lime) &*(lime) &*(lime) &*(lime) &*(lime) & *(purple) & *(purple) & *(purple) & *(purple) & *(purple) & *(purple) \\
*(lime) &*(lime) &*(lime) &*(lime) &*(lime) &*(lime) &*(lime) &*(lime) & *(purple) & *(purple) & *(purple) \\
*(lime) &*(lime) &*(lime) &*(lime) &*(lime) &*(lime) &*(lime) &*(lime)& *(purple) \\
*(lime) &*(lime) &*(lime) &*(lime) &*(lime) &*(lime) &*(lime) &*(lime)\\
*(lime) &*(lime) &*(lime) &*(lime) &*(lime) &*(lime) &*(lime)\\
*(lime) &*(lime) &*(lime) &*(lime) &*(lime) \\
*(lime) &*(lime) 
\end{ytableau}
\end{equation}
Consequently, since the eigenfunctions of the Hamiltonian $\hat{\mathsf{H}}$ form a complete basis of the Hilbert space, the direct sum over the $\mathbf{SU}(n)$ representations corresponding to Young diagrams of the form \eqref{DecompositionL2Gr} yields the decomposition of $L^2\big(\mathsf{Gr}(m,n)\big)$. This is in complete agreement with the spherical weights \eqref{GrassmWeight2}.

\subsubsection{Including monopole.}
Here we need to start by identifying the possible monopole charges ``inside'' Grassmannians, i.e.~the space of invariant 2-forms (which correspond to the field strengths of the monopoles). In the case of Grassmannians, however, it is easier to determine the number of generators of the Picard group, or, in other words, the number of independent inequivalent magnetic line bundles. Since the gauge group \eqref{GLgaugeGroup} here is~$\mathbf{GL}(m,\mathbb{C})$, line bundles can be constructed from the corresponding principal $\mathbf{GL}(m,\mathbb{C})$-bundle and characters of $\mathbf{GL}(m,\mathbb{C})$. In the present case this construction exhausts the Picard group. Indeed, every (algebraic) $\mathbf{GL}(m,\mathbb{C})$-character has the form $g\mapsto(\det g)^k$, where $g\in\mathbf{GL}(m,\mathbb{C})$ and~$k$ is an integer \cite{Fulton1997YoungTableaux}. Thus, the Picard group of the Grassmannian is $\mathbb{Z}$, and the monopole charges are classified by integers, as in the $\mathbb{CP}^{n-1}$ case.

In analogy with Section \ref{SectionCPharmonics}, we can incorporate the magnetic monopole into the first-order GLSM framework by modifying the Grassmannian constraint as
\begin{equation}
VU = 0\qquad\mapsto\qquad VU = \frac{\mathfrak{q}}{2}\,\mathds{1}_m\,,
\end{equation}
where $\mathds{1}_m$ is the $m\times m$ unit matrix. Upon quantization, the modified constraints give the physical-state conditions
\begin{equation}
U_{i\beta}\frac{\partial\psi_{\text{phys}}}{\partial U_{i\alpha}}
=
-\frac{\mathfrak{q}}{2}\,\delta_{\alpha\beta}\psi_{\text{phys}}\,,
\qquad 
\thickbar{U}_{\alpha i}
\frac{\partial\psi_{\text{phys}}}{\partial \thickbar{U}_{\beta i}}
=
\frac{\mathfrak{q}}{2}\,\delta_{\alpha\beta}\psi_{\text{phys}}\,.
\end{equation}

Now we seek a suitable generalization of the ansatz \eqref{AnsatzGr}. To guide the construction, let us first recall the structure of the monopole $\mathbb{CP}^{n-1}$ case \eqref{EigenfunctionsCPq}. There, the wavefunction is constructed by contracting a tensor with two types of building blocks: the gauge-invariant block
\begin{equation}
P^{a\bar b}(u)
=
\frac{u^a\bar{u}^b}{|u|^2}\,,
\qquad
u^k\frac{\partial P^{a\bar b}}{\partial u^k}
=
\bar{u}^k\frac{\partial P^{a\bar b}}{\partial \bar{u}^k}
=
0\,,
\end{equation}
and the ``equivariant'' block
\begin{equation}\label{Mj}
M^{\bar b}(u)
=
\frac{\bar{u}^b}{|u|}\,,
\qquad
u^k\frac{\partial M^{\bar b}}{\partial u^k}
=
-\frac{1}{2}M^{\bar b}\,,
\qquad
\bar{u}^k\frac{\partial M^{\bar b}}{\partial \bar{u}^k}
=
\frac{1}{2}M^{\bar b}\,.
\end{equation}
The monopole charge $\mathfrak{q}$ is then incorporated by taking a product of $\mathfrak{q}$ such blocks\footnote{Here $|u|^{\mathfrak{q}}:=\Big(|u|^2\Big)^{\frac{\mathfrak{q}}{2}}$.},
\begin{equation}
M^{\bar b_1}\ldots M^{\bar b_{\mathfrak{q}}}
=
\frac{\bar{u}^{b_1}\ldots\bar{u}^{b_{\mathfrak{q}}}}{|u|^{\mathfrak{q}}}\,.
\end{equation}

Recall that, in the Grassmannian case, the gauge-invariant block generalizes to rank $m$ projector
\begin{equation}
P^{a\bar b}(U)
=
\Big(U\big(\thickbar{U}U\big)^{-1}\thickbar{U}\Big)^{a\bar b}\,.
\end{equation}
It therefore remains to find an analogue of the elementary equivariant block $M$ with the appropriate transformation properties under the $\mathbf{GL}(m,\mathbb{C})$ gauge group. A natural candidate is
\begin{equation}
M^{\smallthickbar{J}}(U)
=
\frac{\thickbar{\Pi}^{\smallthickbar{J}}}{\sqrt{\det\big(\thickbar{U}U\big)}}\,,
\end{equation}
where $\smallthickbar{J}=\bar b_1\ldots\bar b_m$ is a multi-index and $\Pi$ is the Pl\"ucker coordinate~(\ref{Plucker}). 
The analogue of the equivariance property \eqref{Mj} is\footnote{Here we use the fact that the operator
$U_{k\beta}\frac{\partial}{\partial U_{k\alpha}}$
replaces the index $\alpha$ by $\beta$. In particular, when acting on the determinant, it gives
\begin{equation}
U_{k\beta}\frac{\partial \det\big(\thickbar{U}U\big)}
{\partial U_{k\alpha}}
=
\delta_{\alpha\beta}\det\big(\thickbar{U}U\big)\,.
\end{equation}}
\begin{equation}
U_{k\beta}\frac{\partial M^{\smallthickbar{J}}}{\partial U_{k\alpha}}
=
-\frac{\delta_{\alpha\beta}}{2}M^{\smallthickbar{J}}\,,
\qquad
\thickbar{U}_{\alpha k}
\frac{\partial M^{\smallthickbar{J}}}{\partial \thickbar{U}_{\beta k}}
=
\frac{\delta_{\alpha\beta}}{2}M^{\smallthickbar{J}}\,.
\end{equation}

Thus, the general ansatz for the magnetic wavefunctions on the Grassmannian takes the form
\begin{equation}\label{MagneticAnsatzGr}
\psi_L^{\mathfrak{q}}(U)
=
T_{a_1\ldots a_L|\bar b_1\ldots\bar b_L|\smallthickbar{J}_1\ldots\smallthickbar{J}_{\mathfrak{q}}}\,
P^{a_1\bar b_1}\ldots P^{a_L\bar b_L}
M^{\smallthickbar{J}_1}\ldots M^{\smallthickbar{J}_{\mathfrak{q}}}\,,
\end{equation}
where $T$ is again traceless with respect to the contraction of any holomorphic index with any antiholomorphic index, including those contained in the multi-indices $\smallthickbar{J}_r$.

Under a gauge transformation $U\mapsto Ug$, with $g\in\mathbf{GL}(m,\mathbb{C})$, the wavefunction transforms by the $\mathbf{U}(1)$ phase
\begin{equation}
\psi_L^{\mathfrak{q}}(Ug)
=
\Bigg(\frac{\overline{\det g}}{|\det g|}\Bigg)^{\mathfrak{q}}
\psi_L^{\mathfrak{q}}(U)\,.
\end{equation}
Hence $\psi_L^{\mathfrak{q}}$ can be identified with a section of the line bundle
$\big(\det\mathcal{S}\big)^{\otimes\mathfrak{q}}$, where $\mathcal{S}$ is the tautological bundle\footnote{The tautological bundle is, by definition, the rank-$m$ vector bundle associated with the principal~$\mathbf{GL}(m,\mathbb{C})$ Stiefel bundle
$\mathsf{St}(m,n)\rightarrow\mathsf{Gr}(m,n)$
via the defining representation of $\mathbf{GL}(m,\mathbb{C})$ on $\mathbb{C}^m$.} over $\mathsf{Gr}(m,n)$.

We also need to identify the symmetry type that can be carried by the tensor $T$ in the wavefunctions \eqref{MagneticAnsatzGr} and thus find a suitable generalization of the diagrams \eqref{DecompositionL2Gr}. For this purpose, it is sufficient to understand the role of the Pl\"ucker multi-indices $\smallthickbar{J}$. By construction,~$M^{\smallthickbar{J}}$ carries the $m$-th skew-symmetric $\mathbf{SU}(n)$ tensor representation,
\begin{equation}
M^{\smallthickbar{J}} \qquad\leadsto\qquad \bigwedge^m\thickbar{\mathscr{V}\,}\quad\simeq\quad \bigwedge^{n-m}\mathscr{V}\,.
\end{equation}
The latter isomorphism is given by contraction with the $\varepsilon$-tensor. 
It follows that $M^{\smallthickbar{J}_1}\ldots M^{\smallthickbar{J}_{\mathfrak{q}}}$ belongs a priori to $\mathrm{Sym}^{\text{$\mathfrak{q}$}}\big(\bigwedge^{n-m}\mathscr{V}\big)$. However, the definition \eqref{Plucker}, together with the above $\varepsilon$-tensor identification, realizes $M$ as an exterior product of $n-m$ vectors. Consequently, antisymmetrizing any $n-m+1$ indices in its tensor powers gives zero. This can be equivalently expressed by the Pl\"ucker relations, which in the original antiholomorphic indices read $M^{[\bar a_1\ldots\bar a_m}M^{\bar b_1]\bar b_2\ldots\bar b_m}=0$ (see, for example, \cite{Eastwood1999SomeRO}), where square brackets denote antisymmetrization. Thus any contributing Young diagram has at most $n-m$ rows, giving precisely the $(n-m)\times\mathfrak{q}$ rectangular block.

This leads that the tensor symmetries of \eqref{MagneticAnsatzGr} are the following generalization\footnote{Here we picture the case $n=9$, $m = 4$ and $\mathfrak{q} = 3$.} of~\eqref{DecompositionL2Gr}, namely
\begin{equation}\label{MagneticYoungDiagram}
\begin{ytableau}
*(lime) & *(lime) & *(lime) & *(lime) & *(lime) & *(lime) & *(lime) & *(lime)
& *(orange) & *(orange) & *(orange)
& *(purple) & *(purple) & *(purple) & *(purple) & *(purple) & *(purple) & *(purple) & *(purple) \\
*(lime) & *(lime) & *(lime) & *(lime) & *(lime) & *(lime) & *(lime) & *(lime)
& *(orange) & *(orange) & *(orange)
& *(purple) & *(purple) & *(purple) & *(purple) & *(purple) & *(purple) \\
*(lime) & *(lime) & *(lime) & *(lime) & *(lime) & *(lime) & *(lime) & *(lime)
& *(orange) & *(orange) & *(orange)
& *(purple) & *(purple) & *(purple) \\
*(lime) & *(lime) & *(lime) & *(lime) & *(lime) & *(lime) & *(lime) & *(lime)
& *(orange) & *(orange) & *(orange)
& *(purple) \\
*(lime) & *(lime) & *(lime) & *(lime) & *(lime) & *(lime) & *(lime) & *(lime)
& *(orange) & *(orange) & *(orange) \\
*(lime) & *(lime) & *(lime) & *(lime) & *(lime) & *(lime) & *(lime) \\
*(lime) & *(lime) & *(lime) & *(lime) & *(lime) \\
*(lime) & *(lime)
\end{ytableau}
\end{equation}
Here the \text{\textcolor{LimeGreen}{lime}} boxes represent the indices of ``type $a$'', the \text{\textcolor{purple}{red}} boxes represent the indices of ``type $\bar{b}$'', while the \text{\textcolor{BurntOrange}{orange}} boxes represent the indices of ``type $\smallthickbar{J}\,$'' in \eqref{MagneticAnsatzGr}.

The last step is to determine the magnetic spectrum on the Grassmannian by acting with the Hamiltonian \eqref{HamiltonianGr} on the wave function~\eqref{MagneticAnsatzGr}.
\\
\textbf{Proposition.} The corresponding spectrum is
\begin{empheq}[box = \fcolorbox{black}{white}]{align}
\label{SpectrumGrq}
E^{\text{$\mathsf{Gr}(m,n)$}}_{L_1,\ldots,L_m;\mathfrak{q}}
=
\sum_{j=1}^m
\bigg(L_j+\frac{\mathfrak{q}}{2}\bigg)
\bigg(L_j+\frac{\mathfrak{q}}{2}+n-2j+1\bigg)\,,
\end{empheq}
where $L_1\geq L_2\geq\cdots\geq L_m\geq 0$ are integers labeling the row lengths of the \text{\textcolor{purple}{red}} sub-diagram in the diagram \eqref{MagneticYoungDiagram}.
\begin{proof}
Note that we already know the spectrum in the case $\mathfrak{q}=0$. Thus, it is sufficient to consider the contributions for which at least one derivative in the Hamiltonian \eqref{HamiltonianGr} acts on the monopole blocks $M^{\smallthickbar{J}}$ in the wavefunctions \eqref{MagneticAnsatzGr}.

Let us first determine the energy of the lowest Landau level, which consists of wavefunctions of the form
\begin{equation}
\psi_0^{\mathfrak{q}}
=
T_{\,\smallthickbar{J}_1\ldots\smallthickbar{J}_{\mathfrak{q}}}
M^{\smallthickbar{J}_1}\ldots M^{\smallthickbar{J}_{\mathfrak{q}}}\,.
\end{equation}
A direct, though somewhat tedious, differentiation gives
\begin{equation}\label{ZeroModesEnergy}
\hat{\text{$\mathsf{H}$}}\psi_0^{\mathfrak{q}}
=
-\big(\thickbar{U}U\big)_{\alpha\beta}{\partial}_{\beta i}\thickbar{\partial}_{i\alpha}\psi_0^{\mathfrak{q}}
=
\frac{\mathfrak{q}m}{2}\bigg(n-m+\frac{\mathfrak{q}}{2}\bigg)\psi_0^{\mathfrak{q}}\,.
\end{equation}

Next, we consider the terms in which one derivative acts on a $P$-factor and the other on an $M$-factor. The corresponding cross-terms have the form
\begin{equation}\label{CrossTerms}
-\big(\thickbar{U}U\big)_{\alpha\beta}
\Bigg(
\frac{\partial P^{a\bar b}}{\partial U_{i\beta}}
\frac{\partial M^{\smallthickbar{J}}}{\partial\thickbar{U}_{\alpha i}}
+
\frac{\partial P^{a\bar b}}{\partial\thickbar{U}_{\alpha i}}
\frac{\partial M^{\smallthickbar{J}}}{\partial U_{i\beta}}
\Bigg)
=
P^{a\bar b}M^{\smallthickbar{J}}
+\text{trace terms}\,.
\end{equation}
There are $\mathfrak{q}L$ such terms, where $L=\sum_{j=1}^mL_j$, since there are $L$ factors of $P$ and $\mathfrak{q}$ factors of $M$. The trace terms vanish upon contraction with the traceless tensor $T$. Therefore, the magnetic spectrum is obtained by adding the contributions \eqref{SpectrumGr}, \eqref{ZeroModesEnergy}, and \eqref{CrossTerms}:
\begin{align}
&\sum_{j=1}^m L_j\big(L_j+n-2j+1\big)
+\frac{\mathfrak{q}m}{2}\bigg(n-m+\frac{\mathfrak{q}}{2}\bigg)
+\mathfrak{q}L
=
\nonumber\\
=&\sum_{j=1}^m
\bigg(L_j+\frac{\mathfrak{q}}{2}\bigg)
\bigg(L_j+\frac{\mathfrak{q}}{2}+n-2j+1\bigg)\,.
\end{align}
Thus, the result reproduces \eqref{MagneticCasimirFormula} up to an additive constant in the energy.
\end{proof}

\section{Large-spin limit approach}\label{LargeSpinSection}
In the previous section, we used the quantum-mechanical version of the first-order GLSM formalism to study the spectral problem on unitary Grassmannians $\mathsf{Gr}(m,n)$. In this section, we turn to an alternative approach based on spin chains. From the representation-theoretic point of view, this method allows to recast the space $L^2(\mathcal{M})$ of square-integrable functions on certain homogeneous spaces $\mathcal{M}=\mathbf{G}/\mathbf{H}$ in terms of a tensor product of a finite number of irreducible representations of\,\footnote{Or, possibly, of a central extension $\widehat{\mathbf{G}}$; see \cite{BykovKrivorol2026HolomorphicQuantizations}.} $\mathbf{G}$, in an appropriate large-spin limit. This approach originates in \cite{Bykov:2011ai,Bykov:2012am} and has recently been developed \cite{Bykov:2024tvb,Bykov:2025lcb,Bykov:2024wft,BykovKrivorol2026HolomorphicQuantizations,Krivorol:2026egh}. 

\subsection{Geometric background.}\label{GeometricBAckgroundSubsection}
The basic geometric idea is the following. Suppose that a manifold $\mathcal{M}$ with isometry group $\mathbf{G}$ can be embedded as a Lagrangian submanifold into a product of finitely many coadjoint orbits $\mathcal{O}_k$ of $\mathbf{G}$, i.e.
\begin{equation}
    \mathcal{M}\hookrightarrow
    \bigg(\prod^\times\limits_k\mathcal{O}_k,\sum\limits_k\omega_k\bigg),
    \qquad
    \sum\limits_k\omega_k\Big|_{\mathcal{M}}\equiv 0,
    \qquad
    \dim(\mathcal{M})
    =
    \frac{1}{2}\sum\limits_k\dim\big(\mathcal{O}_k\big),
\end{equation}
where $\omega_k$ is the Kirillov\,--\,Kostant\,--\,Souriau symplectic form on $\mathcal{O}_k$. Then Weinstein's Lagrangian neighborhood theorem states that a neighborhood of $\mathcal{M}$ in the product of coadjoint orbits is symplectomorphic to a neighborhood of the zero section in $\mathrm{T}^\ast\mathcal{M}$ equipped with its canonical symplectic form. In general, however, this local statement is not sufficiently strong for our purposes. There are special situations in which these neighborhoods can be taken to be sufficiently ``large and dense'' so that one may speak, somewhat informally, of an ``almost symplectomorphism''
\begin{equation}
    \big(\mathrm{T}^\ast\mathcal{M},\omega_{\text{can}}\big)
    \quad\text{``}\simeq\text{''}\quad
    \bigg(\prod^\times\limits_k\mathcal{O}_k,\sum\limits_k\omega_k\bigg)\,.
\end{equation}
This stronger relation becomes particularly useful after quantization, since under appropriate conditions it leads to a correspondence between the Hilbert spaces of the two systems,
\begin{equation}\label{GeneralHilbSpaceIsom}
    L^2(\mathcal{M})
    \simeq
    \prod\limits_k^\otimes\mathsf{V}_k\,,
\end{equation}
where $\mathsf{V}_k$ is a representation of $\mathbf{G}$ obtained by quantizing $\big(\mathcal{O}_k,\omega_k\big)$. In situations where the representations $\mathsf{V}_k$ are finite-dimensional, while $L^2(\mathcal{M})$ is infinite-dimensional, this relation should be understood in an appropriate ``large symplectic form'', or large-spin, limit.

Moreover, suppose that $\mathsf{H}_{\text{spin}}$ is a Hamiltonian on the product of coadjoint orbits whose minimum locus is precisely the Lagrangian image of $\mathcal{M}$ and whose Hessian is non-degenerate in the directions transverse to $\mathcal{M}$. Under the symplectic identification with a neighborhood of the zero section in $\mathrm{T}^\ast\mathcal{M}$, the quadratic part of $\mathsf{H}_{\text{spin}}$ in the transverse directions becomes the geodesic Hamiltonian for a metric on $\mathcal{M}$ determined by this Hessian. After quantization, this correspondence leads, in the appropriate large-spin limit, to the Laplace\,--\,Beltrami operator~$-\triangle$ associated with this metric.

\subsection{Unitary Grassmannians revisited.}
For unitary Grassmannians, and more generally for flag manifolds, the corresponding Lagrangian embeddings were constructed in \cite{Bykov:2012am}; see also \cite{AffleckBykov:2021}. In the Grassmannian case, the embedding takes the particularly simple form
\begin{equation}
\mathsf{Gr}(m,n)\hookrightarrow\mathsf{Gr}(n-m,n)\times\mathsf{Gr}(m,n),
\qquad
W\longmapsto\big(W^\perp,W\big)\,,
\end{equation}
where $W\subset\mathbb{C}^n$ is the $m$-dimensional subspace representing a point of $\mathsf{Gr}(m,n)$. The corresponding Hilbert-space isomorphism \eqref{GeneralHilbSpaceIsom} was discussed in \cite{Bykov:2024tvb} in the context of the Laplace\,--\,Beltrami spectral problem. The Lagrangian property of this embedding can be seen directly in Stiefel coordinates. Choose a Stiefel frame $U$ spanning $W$ and orthonormal with respect to the standard Hermitian inner product on $\mathbb{C}^n$, so that $\thickbar{U}U=\mathds{1}_m$, and complete it locally to an orthonormal frame $U_\perp$ spanning $W^\perp$, such that $g=(U,U_\perp)\in\mathbf{U}(n)$. In other words,
\begin{equation}
\thickbar{U}U=\mathds{1}_m\,,
\qquad
\thickbar{U}_\perp U_\perp=\mathds{1}_{n-m}\,,
\qquad
\thickbar{U}U_\perp=0\,,
\qquad
U\thickbar{U}+U_\perp\thickbar{U}_\perp=\mathds{1}_n\,.
\end{equation}
Each of the two Grassmannians carries, up to an overall scale, a unique $\mathbf{U}(n)$-invariant symplectic form. Fixing the common normalization, the pullbacks of the corresponding Kirillov\,--\,Kostant\,--\,Souriau forms are
\begin{equation}\label{Omegas}
\Omega_m=i\,\ell\,\mathrm{Tr}\big(\dif\thickbar{U}\wedge\dif U\big)\,,
\qquad
\Omega_{n-m}=i\,\ell\,\mathrm{Tr}\big(\dif\thickbar{U}_\perp\wedge\dif U_\perp\big)\,,
\end{equation}
where $\ell$ is a constant overall scale of the symplectic forms, which takes positive integer values upon quantization.
The sum of 2-forms \eqref{Omegas} vanishes on the image of the embedding. Indeed,
\begin{equation}
\Omega_m+\Omega_{n-m}
=
i\,\ell\,\mathrm{Tr}\big(\dif\smallthickbar{g}\wedge\dif g\big)
=
-i\,\ell\,\mathrm{Tr}\Big(\big(\smallthickbar{g}\dif g\big)\wedge\big(\smallthickbar{g}\dif g\big)\Big)
=
0\,,
\end{equation}
where in the last step we used $\mathrm{Tr}(A\wedge A)=0$ for a matrix-valued one-form $A$. Finally,
\begin{equation}
\dim_{\mathbb{R}}\mathsf{Gr}(m,n)
=
\frac{1}{2}\dim_{\mathbb{R}}\Big(\mathsf{Gr}(n-m,n)\times\mathsf{Gr}(m,n)\Big)\,,
\end{equation}
and hence the embedding is Lagrangian.

Thus, combining the geometric picture of Section~\ref{GeometricBAckgroundSubsection} with the Borel\,--\,Weil\,--\,Bott theorem, according to which the quantization of $\mathsf{Gr}(k,n)$ yields an irreducible representation corresponding to a rectangular Young diagram of height $k$, we arrive at the essential relation\footnote{This relation was also observed in the context of fuzzy geometry and matrix models in \cite{Dolan:2001mi}; see also~\cite{Huet:2012bk}.}
\begin{equation}\label{L2GrInfiniteLimit}
L^2\big(\text{$\mathsf{Gr}(m,n)$}\big)
=
\lim\limits_{\ell\rightarrow\infty}
\quad
\underbrace
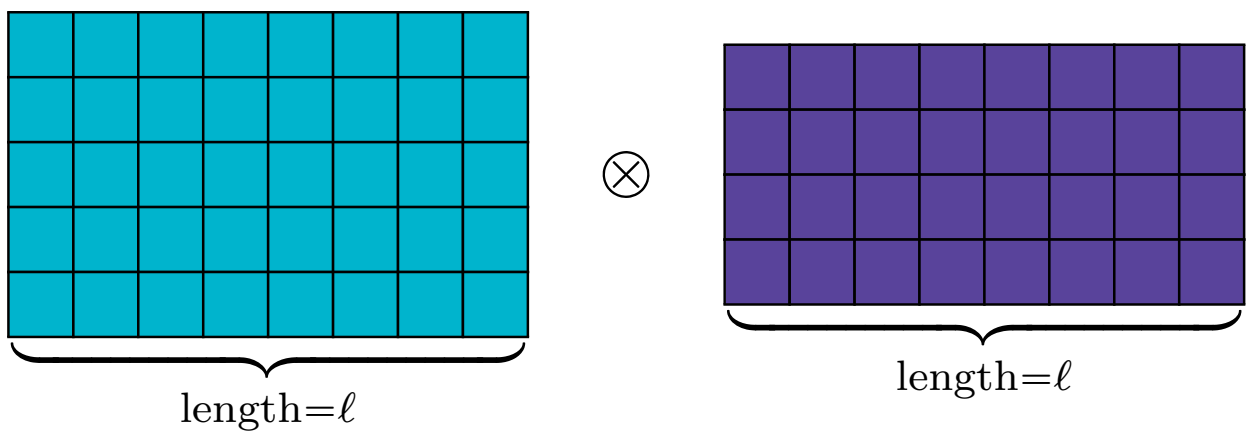
{
\begin{ytableau}
*(Turquoise) & *(Turquoise) & *(Turquoise) & *(Turquoise) & *(Turquoise) & *(Turquoise) & *(Turquoise) & *(Turquoise) \\
*(Turquoise) & *(Turquoise) & *(Turquoise) & *(Turquoise) & *(Turquoise) & *(Turquoise) & *(Turquoise) & *(Turquoise) \\
*(Turquoise) & *(Turquoise) & *(Turquoise) & *(Turquoise) & *(Turquoise) & *(Turquoise) & *(Turquoise) & *(Turquoise) \\
*(Turquoise) & *(Turquoise) & *(Turquoise) & *(Turquoise) & *(Turquoise) & *(Turquoise) & *(Turquoise) & *(Turquoise) \\
*(Turquoise) & *(Turquoise) & *(Turquoise) & *(Turquoise) & *(Turquoise) & *(Turquoise) & *(Turquoise) & *(Turquoise)
\end{ytableau}
}_{\text{length}=\ell}
\quad\otimes\quad
\underbrace{
\begin{ytableau}
*(Violet) & *(Violet) & *(Violet) & *(Violet) & *(Violet) & *(Violet) & *(Violet) & *(Violet) \\
*(Violet) & *(Violet) & *(Violet) & *(Violet) & *(Violet) & *(Violet) & *(Violet) & *(Violet) \\
*(Violet) & *(Violet) & *(Violet) & *(Violet) & *(Violet) & *(Violet) & *(Violet) & *(Violet) \\
*(Violet) & *(Violet) & *(Violet) & *(Violet) & *(Violet) & *(Violet) & *(Violet) & *(Violet)
\end{ytableau}
}_{\text{length}=\ell}\nonumber
\end{equation}
Here\footnote{We again illustrate the case $\mathsf{Gr}(4,9)$.} the first rectangular diagram has height $n-m$, while the second has height $m$. Moreover, parameter $\ell$ of the symplectic form is quantized to a non-negative integer, interpreted as the width of the corresponding diagram. One can check that, in the large-$\ell$ limit, this tensor product reproduces the decomposition into representations corresponding to Young diagrams of the form \eqref{DecompositionL2Gr}. 

We further simplify the analysis by not introducing explicitly the ``spin Hamiltonian''~$\hat{\mathsf{H}}_{\text{spin}}$, as was done in \cite{Bykov:2024tvb}, and instead using the fact that, in the large-spin limit, it reproduces the Laplace\,--\,Beltrami operator. The latter acts on irreducible representations of the symmetry group, as we discussed, through the quadratic Casimir operator, and therefore by a scalar on each irrep, whose value is known from representation theory. Thus, it is sufficient for our purposes to decompose the tensor product above into irreducible representations and compute the corresponding quadratic Casimir eigenvalues.

Let us first fix a finite value of $\ell$. For the tensor product of the two rectangular diagrams under consideration, the Littlewood\,--\,Richardson rule \cite{KoikeTerada1987YoungDiagrammatic} can be organized in the following way \cite[Theorem~2.4]{OKADA1998337}. One decomposes the second rectangle into a Young diagram and its complementary diagram inside the rectangle; schematically,
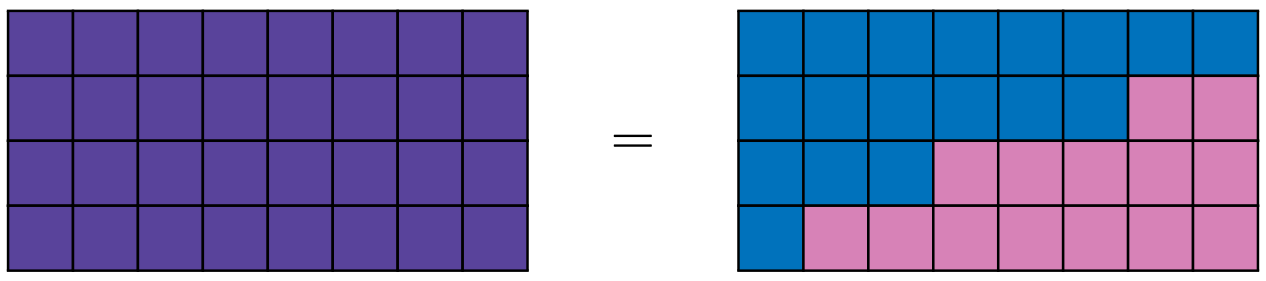
\begin{equation}
\label{SumOverPartitions}
\begin{ytableau}
*(Violet) & *(Violet) & *(Violet) & *(Violet) & *(Violet) & *(Violet) & *(Violet) & *(Violet) \\
*(Violet) & *(Violet) & *(Violet) & *(Violet) & *(Violet) & *(Violet) & *(Violet) & *(Violet) \\
*(Violet) & *(Violet) & *(Violet) & *(Violet) & *(Violet) & *(Violet) & *(Violet) & *(Violet) \\
*(Violet) & *(Violet) & *(Violet) & *(Violet) & *(Violet) & *(Violet) & *(Violet) & *(Violet)
\end{ytableau}
\quad = \quad
\begin{ytableau}
*(RoyalBlue) & *(RoyalBlue) & *(RoyalBlue) & *(RoyalBlue) & *(RoyalBlue) & *(RoyalBlue) & *(RoyalBlue) & *(RoyalBlue) \\
*(RoyalBlue) & *(RoyalBlue) & *(RoyalBlue) & *(RoyalBlue) & *(RoyalBlue) & *(RoyalBlue) & *(Thistle) & *(Thistle) \\
*(RoyalBlue) & *(RoyalBlue) & *(RoyalBlue) & *(Thistle) & *(Thistle) & *(Thistle) & *(Thistle) & *(Thistle) \\
*(RoyalBlue) & *(Thistle) & *(Thistle) & *(Thistle) & *(Thistle) & *(Thistle) & *(Thistle) & *(Thistle)
\end{ytableau}
\end{equation}
The first part is then attached to the right of the first rectangular diagram, while the complementary part is attached below it. This produces
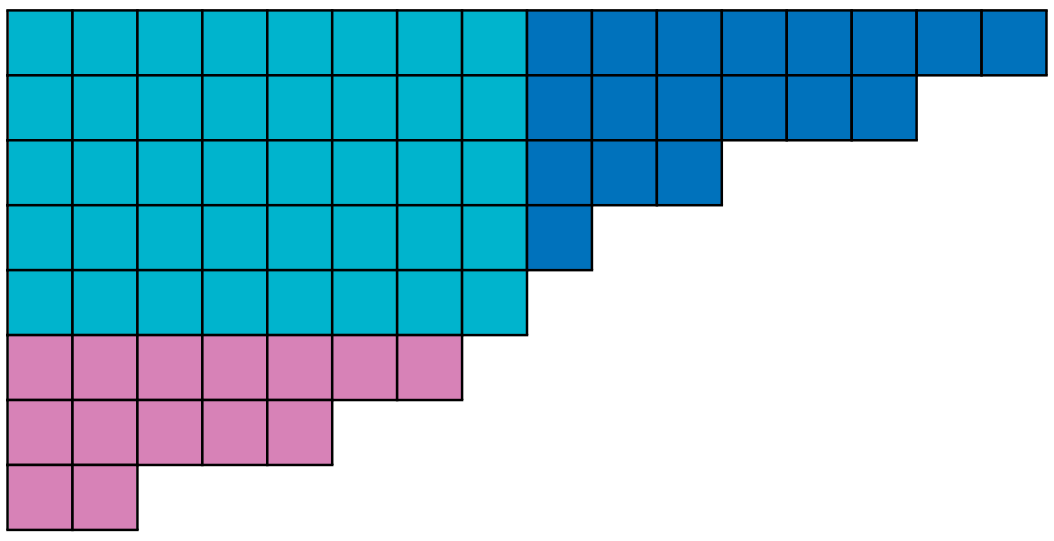
\begin{equation}
\label{FullDiagramsGr}
\begin{ytableau}
*(Turquoise) & *(Turquoise) & *(Turquoise) & *(Turquoise) & *(Turquoise) & *(Turquoise) & *(Turquoise) & *(Turquoise)
& *(RoyalBlue) & *(RoyalBlue) & *(RoyalBlue) & *(RoyalBlue) & *(RoyalBlue) & *(RoyalBlue) & *(RoyalBlue) & *(RoyalBlue) \\
*(Turquoise) & *(Turquoise) & *(Turquoise) & *(Turquoise) & *(Turquoise) & *(Turquoise) & *(Turquoise) & *(Turquoise)
& *(RoyalBlue) & *(RoyalBlue) & *(RoyalBlue) & *(RoyalBlue) & *(RoyalBlue) & *(RoyalBlue) \\
*(Turquoise) & *(Turquoise) & *(Turquoise) & *(Turquoise) & *(Turquoise) & *(Turquoise) & *(Turquoise) & *(Turquoise)
& *(RoyalBlue) & *(RoyalBlue) & *(RoyalBlue) \\
*(Turquoise) & *(Turquoise) & *(Turquoise) & *(Turquoise) & *(Turquoise) & *(Turquoise) & *(Turquoise) & *(Turquoise)
& *(RoyalBlue) \\
*(Turquoise) & *(Turquoise) & *(Turquoise) & *(Turquoise) & *(Turquoise) & *(Turquoise) & *(Turquoise) & *(Turquoise) \\
*(Thistle) & *(Thistle) & *(Thistle) & *(Thistle) & *(Thistle) & *(Thistle) & *(Thistle) \\
*(Thistle) & *(Thistle) & *(Thistle) & *(Thistle) & *(Thistle) \\
*(Thistle) & *(Thistle)
\end{ytableau}
\end{equation}
After forgetting the auxiliary coloring, this is precisely the same Young diagram as the one obtained in \eqref{DecompositionL2Gr}. More explicitly, under the recoloring
\[
\text{\textcolor{Turquoise}{turquoise}},
\ \text{\textcolor{Thistle}{thistle}}
\ \longrightarrow\
\text{\textcolor{LimeGreen}{lime}},
\qquad
\text{\textcolor{RoyalBlue}{blue}}
\ \longrightarrow\
\text{\textcolor{purple}{red}},
\]
the diagram \eqref{FullDiagramsGr} becomes exactly the diagram appearing in \eqref{DecompositionL2Gr}. Thus, the large-spin construction reproduces the same family of irreducible representations as the first-order GLSM approach developed above.

Now, the limit $\ell\rightarrow\infty$ is understood in terms of stabilization of the $\mathbf{SU}(n)$ representation content. For finite $\ell$, the decomposition contains precisely the representations of the type $\mathbb{Y}(L_1,\ldots,L_m)$ with $L_1\leq\ell$, as follows from \eqref{SumOverPartitions}; the corresponding Littlewood\,--\,Richardson diagrams differ from $\mathbb{Y}(L_1,\ldots,L_m)$ only by an overall determinant factor, where $L_i$ are row lengths of the blue sub-diagram in \eqref{FullDiagramsGr}. Hence, every fixed irreducible component appears and stabilizes once $\ell$ is sufficiently large, while the limit $\ell\rightarrow\infty$ simply removes the upper bound on $L_1$.

We now can derive the Grassmannian spectrum by applying the Perelomov\,--Popov \eqref{PerelomovPopov} to the representations  \eqref{FullDiagramsGr}. Let $L_1\geq L_2\geq\cdots\geq L_m\geq0$ denote the row lengths of the \text{\textcolor{RoyalBlue}{blue}} sub-diagram. Since adding a common number of boxes to all $n$ rows only changes the corresponding $\mathbf{U}(n)$ representation by a determinant factor and does not affect the $\mathbf{SU}(n)$ representation, we choose the representative with vanishing last row. Its row lengths are
\begin{equation}\nonumber
\big(R_1,\ldots,R_n\big)
=
\big(
2L_1,L_1+L_2,\ldots,L_1+L_m,
\underbrace{L_1,\ldots,L_1}_{n-2m},
L_1-L_m,L_1-L_{m-1},\ldots,L_1-L_2,0
\big)\,,
\end{equation}
in agreement with \eqref{GrassmannianYoungWeights}. Their average is
\begin{equation}
\frac{1}{n}\sum\limits_{k=1}^n R_k=L_1\,,
\end{equation}
so that the traceless coordinates entering \eqref{PerelomovPopov} are precisely the components of the highest weight \eqref{HighestWeightSU},
\begin{equation}
\big(s_1,\ldots,s_n\big)
=
\big(
L_1,\ldots,L_m,
\underbrace{0,\ldots,0}_{n-2m},
-L_m,\ldots,-L_1
\big)\,.
\end{equation}
Substituting these values into \eqref{PerelomovPopov}, we find
\begin{align}
\mathcal{C}_2(\mathbb{Y})
&=
\frac12\sum\limits_{j=1}^m
\Big[
L_j\big(L_j-2j\big)
+
L_j\big(L_j+2n+2-2j\big)
\Big]
=
\sum\limits_{j=1}^m
L_j\big(L_j+n-2j+1\big)\,,
\label{PerelomovPopovCalculation}
\end{align}
which reproduces exactly the spectrum \eqref{SpectrumGr} obtained from the first-order GLSM approach.

\subsection{Introducing the monopole, again.}

In the previous section, we represented the space of square-integrable functions on the Grassmannian as the large-spin limit of the tensor product of two ``long'' rectangular representations of heights $n-m$ and $m$, respectively, and equal lengths $\ell$. One can naturally generalize this construction by considering rectangles of different lengths, say $\ell^\prime$ and $\ell$, and taking the limit in which both lengths tend to infinity while their difference remains finite. The same stabilization procedure continues to make sense in this case. The difference of the two lengths is then identified with the monopole charge,
$\ell^\prime-\ell=\mathfrak{q}$; see the discussion in \cite{Krivorol:2026egh}. Thus, the corresponding space of square-integrable sections of the magnetic line bundle with monopole charge $\mathfrak{q}$ is represented by the large-spin limit\footnote{Here, for illustration, we picture the case $\mathfrak{q}=3$.}
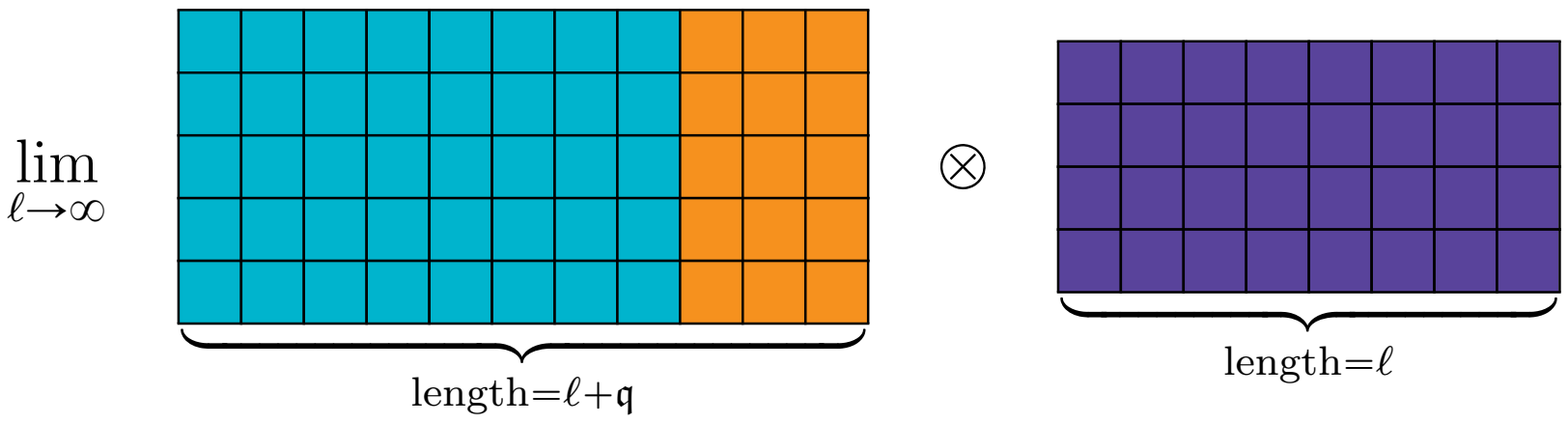
\begin{equation}
\lim\limits_{\ell\rightarrow\infty}
\quad
\underbrace{
\begin{ytableau}
*(Turquoise) & *(Turquoise) & *(Turquoise) & *(Turquoise) & *(Turquoise) & *(Turquoise) & *(Turquoise) & *(Turquoise) & *(BurntOrange) & *(BurntOrange) & *(BurntOrange) \\
*(Turquoise) & *(Turquoise) & *(Turquoise) & *(Turquoise) & *(Turquoise) & *(Turquoise) & *(Turquoise) & *(Turquoise) & *(BurntOrange) & *(BurntOrange) & *(BurntOrange) \\
*(Turquoise) & *(Turquoise) & *(Turquoise) & *(Turquoise) & *(Turquoise) & *(Turquoise) & *(Turquoise) & *(Turquoise) & *(BurntOrange) & *(BurntOrange) & *(BurntOrange) \\
*(Turquoise) & *(Turquoise) & *(Turquoise) & *(Turquoise) & *(Turquoise) & *(Turquoise) & *(Turquoise) & *(Turquoise) & *(BurntOrange) & *(BurntOrange) & *(BurntOrange) \\
*(Turquoise) & *(Turquoise) & *(Turquoise) & *(Turquoise) & *(Turquoise) & *(Turquoise) & *(Turquoise) & *(Turquoise) & *(BurntOrange) & *(BurntOrange) & *(BurntOrange)
\end{ytableau}
}_{\text{length}=\ell+\mathfrak{q}}
\quad\otimes\quad
\underbrace{
\begin{ytableau}
*(Violet) & *(Violet) & *(Violet) & *(Violet) & *(Violet) & *(Violet) & *(Violet) & *(Violet) \\
*(Violet) & *(Violet) & *(Violet) & *(Violet) & *(Violet) & *(Violet) & *(Violet) & *(Violet) \\
*(Violet) & *(Violet) & *(Violet) & *(Violet) & *(Violet) & *(Violet) & *(Violet) & *(Violet) \\
*(Violet) & *(Violet) & *(Violet) & *(Violet) & *(Violet) & *(Violet) & *(Violet) & *(Violet)
\end{ytableau}
}_{\text{length}=\ell}
\end{equation}
so that the first rectangle is longer than the second by $\mathfrak{q}$ \text{\textcolor{BurntOrange}{orange}} columns. As in the non-magnetic case, after removing the irrelevant overall determinant factor, the corresponding tensor product decomposes into $\mathbf{SU}(n)$ representations described by Young diagrams of the form
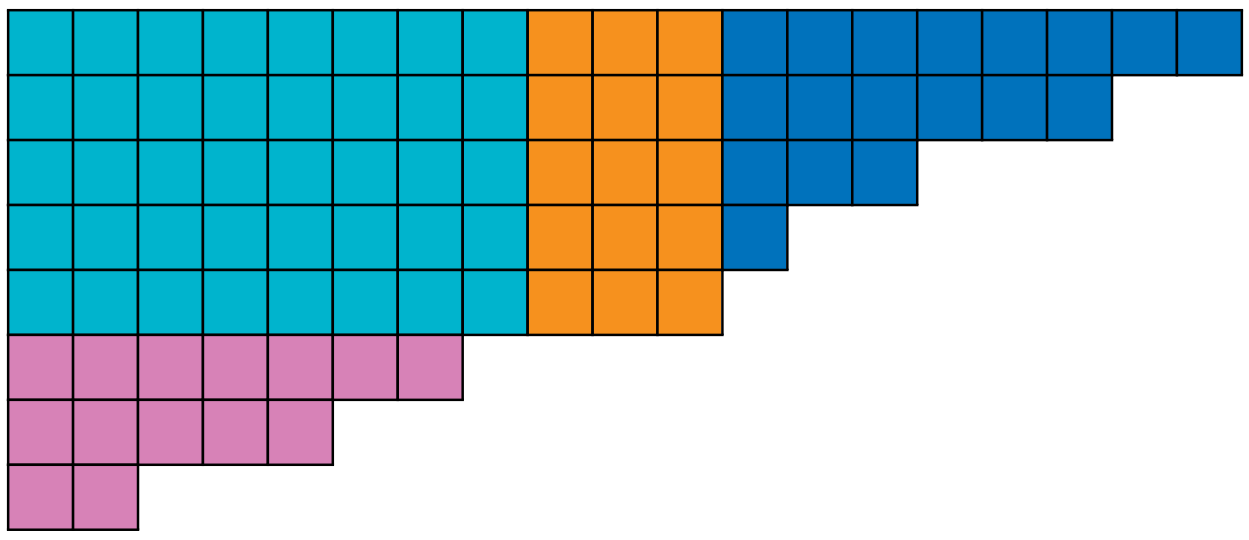
\begin{equation}\label{MagneticYoungDiagramsGr}
\begin{ytableau}
*(Turquoise) & *(Turquoise) & *(Turquoise) & *(Turquoise) & *(Turquoise) & *(Turquoise) & *(Turquoise) & *(Turquoise)
& *(BurntOrange) & *(BurntOrange) & *(BurntOrange)
& *(RoyalBlue) & *(RoyalBlue) & *(RoyalBlue) & *(RoyalBlue) & *(RoyalBlue) & *(RoyalBlue) & *(RoyalBlue) & *(RoyalBlue) \\
*(Turquoise) & *(Turquoise) & *(Turquoise) & *(Turquoise) & *(Turquoise) & *(Turquoise) & *(Turquoise) & *(Turquoise)
& *(BurntOrange) & *(BurntOrange) & *(BurntOrange)
& *(RoyalBlue) & *(RoyalBlue) & *(RoyalBlue) & *(RoyalBlue) & *(RoyalBlue) & *(RoyalBlue) \\
*(Turquoise) & *(Turquoise) & *(Turquoise) & *(Turquoise) & *(Turquoise) & *(Turquoise) & *(Turquoise) & *(Turquoise)
& *(BurntOrange) & *(BurntOrange) & *(BurntOrange)
& *(RoyalBlue) & *(RoyalBlue) & *(RoyalBlue) \\
*(Turquoise) & *(Turquoise) & *(Turquoise) & *(Turquoise) & *(Turquoise) & *(Turquoise) & *(Turquoise) & *(Turquoise)
& *(BurntOrange) & *(BurntOrange) & *(BurntOrange)
& *(RoyalBlue) \\
*(Turquoise) & *(Turquoise) & *(Turquoise) & *(Turquoise) & *(Turquoise) & *(Turquoise) & *(Turquoise) & *(Turquoise)
& *(BurntOrange) & *(BurntOrange) & *(BurntOrange) \\
*(Thistle) & *(Thistle) & *(Thistle) & *(Thistle) & *(Thistle) & *(Thistle) & *(Thistle) \\
*(Thistle) & *(Thistle) & *(Thistle) & *(Thistle) & *(Thistle) \\
*(Thistle) & *(Thistle)
\end{ytableau}
\end{equation}
This decomposition generalizes \eqref{FullDiagramsGr} and reproduces precisely the family of representations described by \eqref{MagneticYoungDiagram}, as expected.

We now extract the spectrum from the quadratic Casimir eigenvalues of the representations \eqref{MagneticYoungDiagramsGr}, in complete analogy with the non-magnetic calculation \eqref{PerelomovPopovCalculation}.
By applying again the Perelomov\,--\,Popov formula \eqref{PerelomovPopov} we obtain
\begin{align}
\mathcal{C}_2^{\mathfrak{q}}(\mathbb{Y})
=
\sum\limits_{k=1}^mL_k\big(L_k+n-2k+1\big)
+
\mathfrak{q}\sum\limits_{k=1}^mL_k
+
\frac{m(n-m)}{2}\mathfrak{q}
+
\frac{m(n-m)}{2n}\mathfrak{q}^2\,.
\end{align}
Equivalently, this result can be written as
\begin{equation}
\mathcal{C}_2^{\mathfrak{q}}(\mathbb{Y}) - \lVert\dif\chi_{\mathfrak{q}}\rVert^2
=
\sum\limits_{j=1}^m
\bigg(L_j+\frac{\mathfrak{q}}{2}\bigg)
\bigg(L_j+\frac{\mathfrak{q}}{2}+n-2j+1\bigg)
-
\frac{m\mathfrak{q}^2}{4}\,,
\end{equation}
where $\chi_{\mathfrak{q}}$ is the isotropy character defining the magnetic line bundle and $\lVert\dif\chi_{\mathfrak{q}}\rVert^2$ is the constant appearing in \eqref{BochnerGrassmannDifference}.
Thus, the spin-chain derivation reproduces the Laplace\,--\,Bochner spectrum obtained above. In particular, comparing with the first-order GLSM spectrum \eqref{SpectrumGrq}, we recover the constant shift $m\mathfrak{q}^2/4$ between the first-order GLSM Hamiltonian and the Laplace\,--\,Bochner operator.

\section{Discussion}
In this work, we consider the spectral problem for the Laplace\,--\,Beltrami and the Laplace\,--\,Bochner operators from several different points of view and using a variety of techniques. We not only review the classical approach to this problem, but also explore more recent ideas arising from the quantization of one-dimensional sigma models in their various formulations. In particular, we employ the so-called first-order GLSM formulation of sigma models, which allows us to cast the corresponding spectral problems in a manifestly $\mathbf{SU}(n)$-equivariant form and to construct the eigenfunctions in a particularly simple and geometrically transparent way, in terms of Pl\"ucker coordinates and Grassmannian projectors. Furthermore, using the relation between sigma models and ``short'' spin chains, we obtain a particularly simple organization of the representation-theoretic structure of Grassmannian harmonics, recasting the corresponding Peter\,--\,Weyl decomposition in terms of the large-spin limit of a tensor product of two rectangular $\mathbf{SU}(n)$ representations.

There are also several interesting directions for future research. One natural extension would be to generalize our sigma model approaches to isotropic Grassmannians, i.e.~orthogonal and symplectic Grassmannians, or even to isotropic flag manifolds. These spaces form a prototypical family containing both symmetric and non-symmetric examples, and therefore provide a natural testing ground for understanding how the representation-theoretic complexity of the non-symmetric cases simplifies in the symmetric ones. The corresponding first-order GLSM formulation for isotropic Grassmannians was constructed in \cite{BykovKrivorol2023}, while the first steps toward developing the spin-chain approach were taken in \cite{Bykov:2025wvn}. It is also interesting to include supersymmetry in this picture. The ``worldline'' $\mathcal{N}=2$ and $\mathcal{N}=4$ supersymmetric sigma models describe the spectral problems for the Laplace\,--\,Dolbeault and Laplace\,--\,de Rham operators, respectively. The corresponding spin-chain formulations were constructed in \cite{Bykov:2024wft}, while the first-order GLSM formulations were developed in \cite{BykovKrivorolSUSY}. Another interesting direction is provided by ``target-space'' supersymmetry, i.e.~sigma models on supergroup cosets \cite{Ivanov:2004yw,Murray:2006pi,Chuah:2024eii, TanSuperSigmaModels}.

Another class of manifolds for which our methods may prove useful consists of total spaces of bundles over Grassmannians, and more generally over flag manifolds. A natural family of examples is provided by Stiefel manifolds; see \cite{AgricolaCaglieroHenkel2026ExplicitLaplace,BirteaCasuComanescu2026Stiefel,Strichartz1975ExplicitFourier,Gelbart1974StiefelHarmonics}. Quaternionic projective spaces (cf.~\cite{Grantcharov:2021han})  provide another natural class of examples to which our methods may be applied.

\subsection*{Acknowledgments}
Viacheslav Krivorol thanks Mikhail Markov for useful discussions. 
The work of Dmitri Bykov was performed within the framework of the state assignment at Steklov Mathematical Institute of Russian Academy of Sciences. 
The work of Viacheslav Krivorol was supported by the Russian Science Foundation grant №25-72-10177.

\setstretch{0.8}
\setlength\bibitemsep{5pt}
\printbibliography
\end{document}